\documentclass[12pt]{iopart}

\usepackage{graphicx}
\usepackage{bm}
\usepackage{psfrag}
\usepackage{iopams}

\begin{document}

\title[]{The probability distribution of internal stresses in externally loaded 2D dislocation systems}

\author{P\'eter Dus\'an Isp\'anovity and Istv\'an Groma}

\address{Department of Materials Physics, E\"otv\"os University, P\'azm\'any P\'eter s\'et\'any 1/A, Budapest H-1117, Hungary}

\ead{ispanovity@metal.elte.hu, groma@metal.elte.hu}

\begin{abstract}
The distribution of internal shear stresses in a 2D dislocation system is investigated when external shear stress is applied. This problem serves as a natural continuation of the previous work of Csikor and Groma (Csikor F F and Groma I 2004 \textit{Phys.\ Rev.}\ B \textbf{58} 2969), where analytical result was given for the stress distribution function at zero applied stress. First, the internal stress distribution generated by a set of randomly positioned ideal dislocation dipoles is studied. Analytical calculations are carried out for this case. The theoretical predictions are checked by numerical simulations showing perfect agreement. It is found that for real relaxed dislocation configurations the role of dislocation multipoles cannot be neglected, but the theory presented can still be applied.
\end{abstract}



\submitto{Journal of Statistical Mechanics: Theory and Experiment}

\maketitle

\section{Introduction}

The macroscopic plastic deformation of a crystalline material is the result of the movement of a huge number of dislocations interacting with each other through the long-range anisotropic shear stress field generated by a single dislocation. This feature and the fact that dislocation movement is often constrained to single glide plane result in dynamics exhibiting very high spatiotemporal complexity. This is indicated, for instance, by the multitude of observed dislocation patterns or by the recently revealed fact that plastic deformation is characterised by intermittent strain bursts with scale-free size distribution \cite{miguel, dimiduk, richeton, csikor_science}.

Statistical descriptions of dislocation systems mainly concentrate on deriving continuum models for the evolution of the different dislocation densities. Several such theories have been proposed so far both for two \cite{sedlacek, schwarz, groma1, groma2, groma3} and three dimensions \cite{elazab1, elazab2, zaiser1, hochrainer, zaiser2}. In all these models a key problem is to take into account dislocation-dislocation correlations during the coarse graining of the microstructure. (It is known that neglecting the dislocation-dislocation correlations leads to unphysical results \cite{groma2, groma3}.) It is found that correlation effects enter into the theories through gradient-like stress terms \cite{schwarz, groma2, zaiser2}, which were already proposed earlier phenomenologically \cite{walgraef1, walgraef2}. Motivated by its crucial role many investigations on the properties of dislocation correlations have been published recently \cite{zaisermiguel, deng, csikor_corr, vinogradov, ispanovity}.

Beside the one-particle dislocation density function the probability distribution function of internal stresses serves as another important quantity for the statistical characterisation of dislocation systems. Previously this function was determined for relaxed 2D dislocation configurations \cite{gromabako, csikor} as well as for fractal-like dislocation morphologies \cite{beato}, but in these studies the role of external shear stress was not investigated. For many aspects, however, the external stress obviously plays a major role.

What gives the stress distribution function outstanding importance is the fact that it can be directly measured on real materials by X-ray diffraction, even in situ. Briefly, it was shown by Groma and co-workers that the shape of the broadened Bragg peaks depend on the distribution of the internal strain \cite{groma4, szekely, borbely} that is proportional to the stress. Consequently, from the Bragg peaks the stress distribution function can be extracted.

Another practical issue is the numerical modelling of discrete dislocations. It was observed by Groma and Bak\'o that the force acting on a dislocation can be divided into a slowly varying component coming from the further dislocations and a stochastic contribution of the near dislocations \cite{bakogroma}. In the proposed $\mathcal{O}(N)$ stochastic dislocation dynamics simulation method the latter is drawn from the stress distribution function \cite{gromabako, groma5}.

Like in the previous studies on the topic \cite{gromabako, csikor, beato} we consider two dimensional (2D) dislocation systems. This makes the calculations much simpler, but according to X-ray studies, under quite general conditions the results are applicable for deformed 3D crystals as well. On the other hand, there is a growing interest towards recently observed physical systems with real 2D underlying lattices whose macroscopic properties are highly influenced by dislocations. Examples include dusty plasma crystals where particles are arranged in a hexagonal lattice \cite{quinn, nosenko}, vortex lattices in type II superconducting films \cite{miguel2}, colloidal crystals \cite{murray, schall} and foams \cite{kader}.

In this paper the stress distribution function is investigated in the presence of external loading. In section \ref{sec:analytic_form}, theoretical calculations are carried out to arrive at a closed analytical expression for the Fourier transform of the distribution function. This expression depends on the spatial dislocation-dislocation correlation function. Analytical expression has not been found yet for the correlation function, it can be determined only numerically \cite{zaisermiguel, ispanovity}. To arrive at close form for the Fourier transform of the stress distribution function, like in an earlier study \cite{csikor}, in section \ref{sec:stress} we consider randomly distributed monodisperse dislocation dipole systems. This permits to give the asymptotic decay of the stress distribution (section \ref{sec:evolv}). It is found that due to external load the stress distribution function becomes asymmetric. Namely, a $1/(\tau |\tau|^3)$ like term proportional with the applied external stress is added to the $1/|\tau|^3$ like tail of the distribution function ($\tau$ denotes the shear stress). In the vicinity of the origin the distribution function is shifted with a value which is again proportional to the applied external stress.

Stress distribution functions obtained numerically are presented in section \ref{sec:numerics}. First, the theoretical predictions are validated on the monodisperse dipole systems. Then investigations of relaxed configurations obtained by discrete dislocation dynamics simulations follow. It is found that the asymptotic decay seen on monodisperse systems remains valid. Difference is found only in the numerical values of some parameters, which is attributed to the large number of dislocation multipoles present in the relaxed configurations making the system more resistant against the external stress.

\section{The analytical form of the stress distribution function}
\label{sec:analytic_form}

The general mathematical formulation of the stress distribution function was given previously by Groma and Bak\'o \cite{gromabako}. In this section, first, we briefly present their results then continue with analysis specific to the problem subject of this paper.

\subsection{General description}

Let us consider a system of $N$ parallel straight edge dislocations with line directions parallel to the $z$ axis. For the sake of simplicity we restrict our analysis for single glide, so the Burgers vector of the dislocations can be only $\pm \bm b$ with $\bm b$ taken to be parallel to the $x$ axis. Under these conditions, the dislocations' movement is parallel with the $x$ axis and the three dimensional problem can be treated as a 2D one in the $xy$ plane. The position and the sign of the Burgers vector of the \textit{i}th dislocation in the $xy$ plane is denoted by $\bm r_i$, and $s_i$ ($\bm r_i \in \mathbb{R}^2$, $s_i \in \{-1,1\}$ and $i \in \{1, \dots, N\}$), respectively.

Assuming linear elasticity the internal shear stress field generated by the $N$ dislocations at a given point $\bm r \in \mathbb R^2$ is
\begin{eqnarray}
	\tau_\mathrm{str}^{s_1,\dots, s_N}(\bm r, \bm r_1, \dots, \bm r_N) :=
	\sum_{i=1}^N s_i \tau_{\mathrm{ind}}(\bm r - \bm r_i),
	\label{eqn:stress_r}
\end{eqnarray}
where $\tau_{\mathrm{ind}}$ is the shear stress field in the $xy$ plane generated by an individual positive sign edge dislocation positioned in the origin:
\begin{eqnarray}
\tau_{\mathrm{ind}}(\bm r) := Gb\frac{x(x^2-y^2)}{(x^2+y^2)^2} = Gb \frac{\sin(\varphi) \cos(2 \varphi)}{r},
\label{eqn:tauind}
\end{eqnarray}
in which $G$ is a combination of the elastic moduli: $G=\frac{\mu}{2\pi(1-\nu)}$, where $\mu$ and $\nu$ are the shear modulus and Poisson's ratio, respectively. It has to be noted, that the given form of $\tau_\mathrm{ind}$ is valid only for dislocations in an infinite medium.

The subject of this paper is to determine the stress distribution function $P_{\mathrm{str}}$ at a randomly chosen point $\bm r$. By definition $P_{\mathrm{str}}(\tau, \bm r) \rmd \tau$ is the probability of finding the stress generated by the system between the limits
\begin{eqnarray}
\tau - \frac{\rmd \tau}{2} < \tau_\mathrm{str}^{s_1, \dots, s_N}(\bi r, \bm r_1, \dots, \bm r_N) < \tau + \frac{\rmd \tau}{2}.
\end{eqnarray}
To determine $P_{\mathrm{str}}$ Markoff's method \cite{markoff} is followed.

Let $w_N^{s_1, \dots, s_N}$ denote the $N$ particle distribution function of the system. After introducing the function
\begin{eqnarray}
\fl \Delta^{s_1, \dots, s_N}(\bm r, \bm r_1, \dots, \bm r_N, \tau, \rmd \tau) :=
	\left\{
	\begin{array}{ll}
	 1, & \mbox{ if $\tau - \frac{\rmd \tau}{2} < \tau_\mathrm{str}^{s_1, \dots, s_N}(\bm r, \bm r_1, \dots, \bm r_N) < \tau + \frac{\rmd \tau}{2}$,} \\
	 0, & \mbox{ otherwise}
	\end{array}
	\right., \nonumber \\
\end{eqnarray}
$P_{\mathrm{str}}$ can be expressed in the form
\begin{eqnarray}
\fl P_{\mathrm{str}}(\tau, \bm r) \rmd \tau &= \sum_{s_1=\pm 1} \dots \sum_{s_N= \pm 1} \int\limits_{\mathbb R^2} \rmd^2 r_1 \dots \int\limits_{\mathbb R^2} \rmd^2 r_N \Delta^{s_1, \dots, s_N}(\bm r, \bm r_1, \dots, \bm r_N, \tau, \rmd \tau) \nonumber \\
\fl & \quad \times w_N^{s_1, \dots, s_N}(\bm{r}_1, \dots, \bm{r}_N). \label{eqn:pvegtelen}
\end{eqnarray}
It was previously pointed out by Groma and Bak\'o that, in contrast to Markoff's original deduction, the dislocation-dislocation correlations must be taken into account to avoid the system size dependence of the stress distribution function \cite{gromabako}. Then the resulting form of the Fourier transform of the probability density function ($P_\mathrm{str}^F(q, \bm r):=\frac{1}{2\pi} \int_{\mathbb R} P_\mathrm{str}(\tau, \bm r) \rme^{-\rmi q\tau} \rmd \tau$) is
\begin{eqnarray}
\fl \ln \left( P_\mathrm{str}^F(q, \bm r) \right) =& -\sum_{s_1=\pm 1} \int\limits_{\mathbb R^2} \rho_1^{s_1}(\bm r_1)B^{s_1}(\bm r - \bm r_1, q) \rmd^2 r_1 + \frac{1}{2} \sum_{s_1=\pm 1} \sum_{s_2= \pm 1} \int\limits_{\mathbb R^2}\int\limits_{\mathbb R^2} \rho_1^{s_1}(\bm r_1) \rho_1^{s_2}(\bm r_2) \nonumber \\
\fl & \times d_2^{s_1, s_2}(\bm r_1, \bm r_2) B^{s_1}(\bm r - \bm r_1, q) B^{s_2}(\bm r - \bm r_2, q) \rmd^2 r_1 \rmd^2 r_2 + \dots, \label{eqn:lnP^F}
\end{eqnarray}
where
\begin{eqnarray}
B^s(\bm r, q) := 1 - \exp \bm ( \rmi q s \tau_{\mathrm{ind}}(\bm r) \bm ),
\label{eqn:B^s}
\end{eqnarray}
and $d_2^{s_1, s_2}$ denotes the dislocation-dislocation correlation functions defined as
\begin{eqnarray}
d_2^{s_1, s_2}(\bm r_1, \bm r_2) := \frac{\rho_2^{s_1, s_2}(\bm r_1, \bm r_2)}{\rho_1^{s_1}(\bm r_1)\rho_1^{s_2}(\bm r_2)} - 1
\label{eqn:d_2}
\end{eqnarray}
in which $\rho_1^{s}$ and $\rho_2^{s_1, s_2}$ are the one and the two particle dislocation density functions, respectively. The superscripts in these functions refer to the sign of the dislocations, e.g., $\rho_2^{+-}(\bm r_1, \bm r_2)$ is proportional to the probability of finding a positive dislocation at $\bm r_1$ and a negative one at $\bm r_2$.

In order to evaluate (\ref{eqn:lnP^F}), further assumptions have to be made. In the rest of this paper we consider only neutral infinite homogeneous configurations, meaning $\rho_1^s$ is constant in space:
\begin{eqnarray}
\rho^s_{\mathrm{dis}}:=\rho_1^s(\bm r) \label{eqn:const_rho}
\end{eqnarray}
and the densities of the positive and negative sign dislocations are equal:
\begin{eqnarray}
\rho_{\mathrm{dis}}:=2\rho_{\mathrm{dis}}^+=2\rho_{\mathrm{dis}}^-.
\end{eqnarray}
Consequently,
\begin{itemize}
\item The direct $\bm r$ dependence of $P_\mathrm{str}$ must vanish and
\item The $d_2^{s_1, s_2}$ correlation functions depend only on the difference of their arguments \cite{zaisermiguel, ispanovity}:
\begin{eqnarray}
d_2^{s_1, s_2}(\bm r_1, \bm r_2) = d_2^{s_1, s_2}(\bm r_1 - \bm r_2).
\end{eqnarray}
\end{itemize}
We note that for symmetry reasons $d_2^{--}(\bm r) = d_2^{++}(\bm r)$ and $d_2^{-+}(\bm r) = d_2^{+-}(-\bm r)$ hold for all $\bm r$ values, thus the correlations can be described using only the $d_2^{++}$ and the $d_2^{+-}$ functions \cite{zaisermiguel, ispanovity}. As in previous studies, we neglect the three and higher order correlations meaning that the terms not written out explicitly in (\ref{eqn:lnP^F}) are omitted.

\subsection{The symmetric part of the distribution's Fourier transform}

In the absence of external stress the dislocation pair correlation functions are symmetric: $d_2^{s_1,s_2}(\bm r) = d_2^{s_1,s_2}(-\bm r)$ with $s_1, s_2 \in \{+,-\}$ \cite{zaisermiguel, ispanovity}. Then, according to (\ref{eqn:lnP^F}), $P_\mathrm{str}^{F}$ is real, and therefore the distribution function $P_\mathrm{str}$ is symmetric. This case was thoroughly studied recently by Csikor and Groma \cite{csikor}. Starting from (\ref{eqn:lnP^F}), for relaxed dislocation configurations they found asymptotes for $P_\mathrm{str}^F$ in the limits of $|q| \to 0$ and $|q| \to \infty$.

Under applied external shear stress the correlation functions of the system become different (for details see section \ref{sec:corr}) which modifies the second term in (\ref{eqn:lnP^F}). After a short calculation one concludes that the real part of (\ref{eqn:lnP^F}) depends only on  $d_2^{++}$ and the combination $d_2^{+-} + d_2^{-+}$. Since these functions do not change considerably due to the external stress (see section \ref{sec:corr} for details), we assume that $\mathrm{Re}\left( P_\mathrm{str}^F \right)$ can be well approximated by the form derived by Csikor and Groma \cite{csikor}:
\begin{eqnarray}
\mathrm{Re} \left[ \ln\left(P_{\mathrm{str}}^{F}(q)\right) \right]=\left\{
\begin{array}{ll}
\displaystyle C \rho_{\mathrm{dis}} q^2 \ln \left( \frac{|q|}{q_{\mathrm{eff}}} \right), & \mathrm{if} \  |q| \to 0, \\
\\
\displaystyle -\frac{D}{2} \rho_{\mathrm{dis}}|q|, & \mathrm{if} \  |q| \to \infty,
\end{array}
\right.
\label{eqn:re_P}
\end{eqnarray}
where $C=\frac{\pi}{4}(Gb)^2$, $D$ is a parameter which can only be determined numerically (for relaxed systems $D = 1.35\,Gb\rho_\mathrm{dis}^{-0.5}$ was obtained) and $q_\mathrm{eff}$ is a constant \cite{csikor} but it does not play role in the further considerations.

\subsection{The antisymmetric part of the distribution's Fourier transform}
\label{sec:antisymm}

After substituting $B^s$ defined by (\ref{eqn:B^s}) into (\ref{eqn:lnP^F}) and performing a few straightforward transformations one obtains
\begin{eqnarray}
\mathrm{Im} \left[ \ln\left(P_{\mathrm{str}}^{F}(q)\right) \right] = \left( \rho_{\mathrm{dis}}/2 \right)^2 \int\limits_{\mathbb R^2} d_2^{+-}(\bm d)T(\bm d, q) \rmd^2 d, \label{eqn:im_P}
\end{eqnarray}
where
\begin{eqnarray}
T(\bm d, q) := \int\limits_{\mathbb R^2} \sin \left( q \left[ \tau_{\mathrm{ind}} \left( \bm r - \frac {\bm d}2 \right)-\tau_{\mathrm{ind}} \left( \bm r + \frac {\bm d}2 \right) \right] \right) \rmd^2 r \label{eqn:T(d,q)},
\end{eqnarray}
in which it is assumed that $d_2^{++}(\bm r) = d_2^{++}(-\bm r)$, and so the second term in (\ref{eqn:lnP^F}) vanishes for $s_1=s_2=1$ and $s_1=s_2=-1$. This assumption will be verified in section \ref{sec:corr}.

The main subject of this paper is the evaluation of (\ref{eqn:im_P}). If no external stress is applied, the right hand side of (\ref{eqn:im_P}) vanishes \cite{csikor}, but in a loaded system this term plays an important role. In the following, first, the behaviour of function $T$ is investigated by performing the integral in (\ref{eqn:T(d,q)}) numerically. Afterwards, the properties of $\mathrm{Im} \left[ \ln\left(P_{\mathrm{str}}^{F}(q)\right) \right]$ are analysed in detail.

Since analytical solution for the integral appearing in (\ref{eqn:T(d,q)}) was not found, it is calculated numerically. Because of the unusual form of the integral, however, the numerical integration is not trivial. First, we have to make an important remark.

Integration in (\ref{eqn:T(d,q)}) should be performed on the whole $\mathbb R^2$ plane, but with any numerical algorithm one can calculate the integral for only a finite region and study its region size dependence. Formally speaking, the integral of an arbitrary $f:\mathbb R^2 \to \mathbb R$ function is approximated as:
\begin{eqnarray}
\int\limits_{\mathbb R^2} f(\bm r) \rmd^2 r = \lim_{R \to \infty} \int\limits_0^R \rmd r \, r \int\limits_{[0, 2 \pi[} \rmd \varphi \, f(r, \varphi).
\end{eqnarray}
This approximation inherently assumes that the $f$ function can be integrated successively (first for $\varphi$ then for $r$) which is not true for all possible integrable $f$ functions. According to Fubini's theorem, for successive integration, among others,
\begin{eqnarray}
\int\limits_0^\infty r | f(r, \varphi) | \rmd r < + \infty \label{eqn:fubini}
\end{eqnarray}
must hold \cite{rudin}.

Since $\tau_\mathrm{ind}$ exhibits a $r^{-1}$ type decay, the argument of the sine function in (\ref{eqn:T(d,q)}) has a $r^{-2}$ type asymptote as $r \rightarrow \infty$. This implies that for the first term of the expansion of the sine function in (\ref{eqn:T(d,q)}), which also has a $r^{-2}$ tail, the condition given by (\ref{eqn:fubini}) is not fulfilled. So, according to the aforementioned theorem, the integral in (\ref{eqn:T(d,q)}) cannot be performed successively, and therefore it cannot be determined by a straightforward numerical algorithm:
\begin{eqnarray}
T(\bm d, q) & = \int\limits_{\mathbb R^2} \sin \left( q \left[ \tau_{\mathrm{ind}} \left( \bm r - \frac {\bm d}2 \right)-\tau_{\mathrm{ind}} \left( \bm r + \frac {\bm d}2 \right) \right] \right) \rmd^2 r \nonumber \\
& \ne \int\limits_0^\infty \rmd r \, r \int\limits_0^{2 \pi} \rmd \varphi \sin \left( q \left[ \tau_{\mathrm{ind}} \left( \bm r - \frac {\bm d}2 \right)-\tau_{\mathrm{ind}} \left( \bm r + \frac {\bm d}2 \right) \right] \right).
\end{eqnarray}
In order to overcome this problem we have to note that
\begin{itemize}
\item The higher order terms in the expansion of the sine function in (\ref{eqn:T(d,q)}) have $r^{-6}$, $r^{-10}$, \dots~type singularities and so can be integrated successively.
\item The integral of the first term [which breaks the condition (\ref{eqn:fubini})] must be zero, because $\tau_{\mathrm{ind}}$ is an odd function, and hence
\begin{eqnarray}
& \int\limits_{\mathbb R^2} q \left[ \tau_{\mathrm{ind}} \left( \bm r - \frac {\bm d}2 \right)-\tau_{\mathrm{ind}} \left( \bm r + \frac {\bm d}2 \right) \right] \rmd^2 r \nonumber \\
&\quad = q \int\limits_{\mathbb R^2} \tau_{\mathrm{ind}} \left( \bm r - \frac {\bm d}2 \right) \rmd^2 r - q \int\limits_{\mathbb R^2} \tau_{\mathrm{ind}} \left( \bm r + \frac {\bm d}2 \right) \rmd^2 r = 0.
\end{eqnarray}
\end{itemize}
Accordingly, if the first term of the series of the sine function is subtracted from the argument of the integral in (\ref{eqn:T(d,q)}), then what remains can already be integrated successively. So,
\begin{eqnarray}
T(\bm d, q) &= \int\limits_{\mathbb R^2} \sin \left( q \left[ \tau_{\mathrm{ind}} \left( \bm r + \frac {\bm d}2 \right)-\tau_{\mathrm{ind}} \left( \bm r - \frac {\bm d}2 \right) \right] \right) \rmd^2 r \nonumber \\
&= \int\limits_0^\infty \int\limits_0^{2 \pi} \left\{ \sin \left( q \left[ \tau_{\mathrm{ind}} \left( \bm r + \frac {\bm d}2 \right)-\tau_{\mathrm{ind}} \left( \bm r - \frac {\bm d}2 \right) \right] \right) \right. \nonumber \\
& \quad \left. - q \left[ \tau_{\mathrm{ind}} \left( \bm r + \frac {\bm d}2 \right)-\tau_{\mathrm{ind}} \left( \bm r - \frac {\bm d}2 \right) \right] \right\}
 r \, \rmd \varphi \, \rmd r.
\label{eqn:T}
\end{eqnarray}

\begin{figure}[!ht]
\begin{center}
\hspace*{-0.7cm}
\includegraphics[width=6cm,angle=-90]{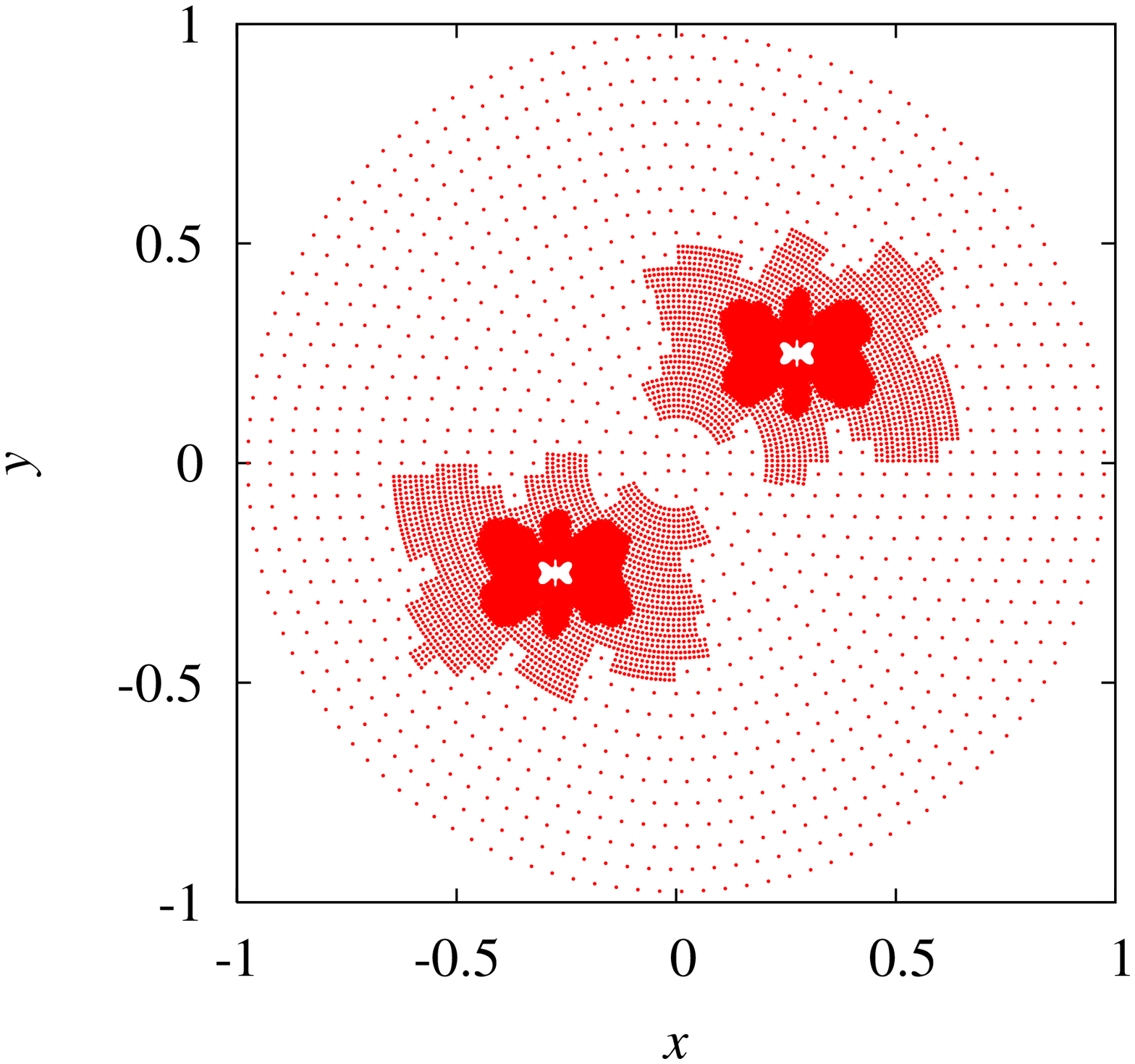}
\caption{\label{fig:basepoints} The grid-points of the algorithm developed for evaluating the integral in (\ref{eqn:T}) next to the points $\bm d=\left( \begin{array}{cc} 0.55\\ 0.5 \end{array} \right) $. Closer grid-points are taken near the singular points $\bm r = \pm \bm d/2$.}
\end{center}
\end{figure}

With this, the determination of the function $T$ is now simplified to the numerical integration of the right hand side of (\ref{eqn:T}). During the numerical computation, however, special attention is needed at the points $\bm r = \pm \frac {\bm d}2$ since the argument of the sine function has a $1/r$ type singularities there. Moreover, the argument exhibits a strong angular dependence around $\pm \frac {\bm d}2$, too. As a result, the first term of the integrand in (\ref{eqn:T}) oscillates rapidly between $1$ and $-1$ near the mentioned points. To handle this behaviour correctly, close to the points $\bm r = \pm \frac {\bm d}2$ the integration grid-point distance was adaptively reduced. Formally, denoting by $\Delta x$ the distance of the neighbouring grid-points at a given grid-point if the condition
\begin{eqnarray}
\left| q \Delta x \cdot \nabla_{\bm r} \left[ \tau_{\mathrm{ind}} \left( \bm r + \frac {\bm d}2 \right) - \tau_{\mathrm{ind}} \left( \bm r - \frac {\bm d}2 \right) \right] \right| < \frac{2\pi}{N_\mathrm{b}}
\label{eqn:trapez}
\end{eqnarray}
is not fulfilled additional grid-points are introduced. [\Eref{eqn:trapez} corresponds to the criterion that there must be at least $N_\mathrm{b}$ grid-points in every period of the oscillating function.] The refinement of the mesh has to be stopped at a certain distance from the singular points. However, it is easy to see that the integral for a small symmetric domain around $\pm \frac {\bm d}2$ vanishes as the size of that domain approaches zero. The mesh used is demonstrated in figure \ref{fig:basepoints}.

\begin{figure}[!ht]
\begin{center}
\hspace*{-1.5cm}
\includegraphics[width=6cm,angle=-90]{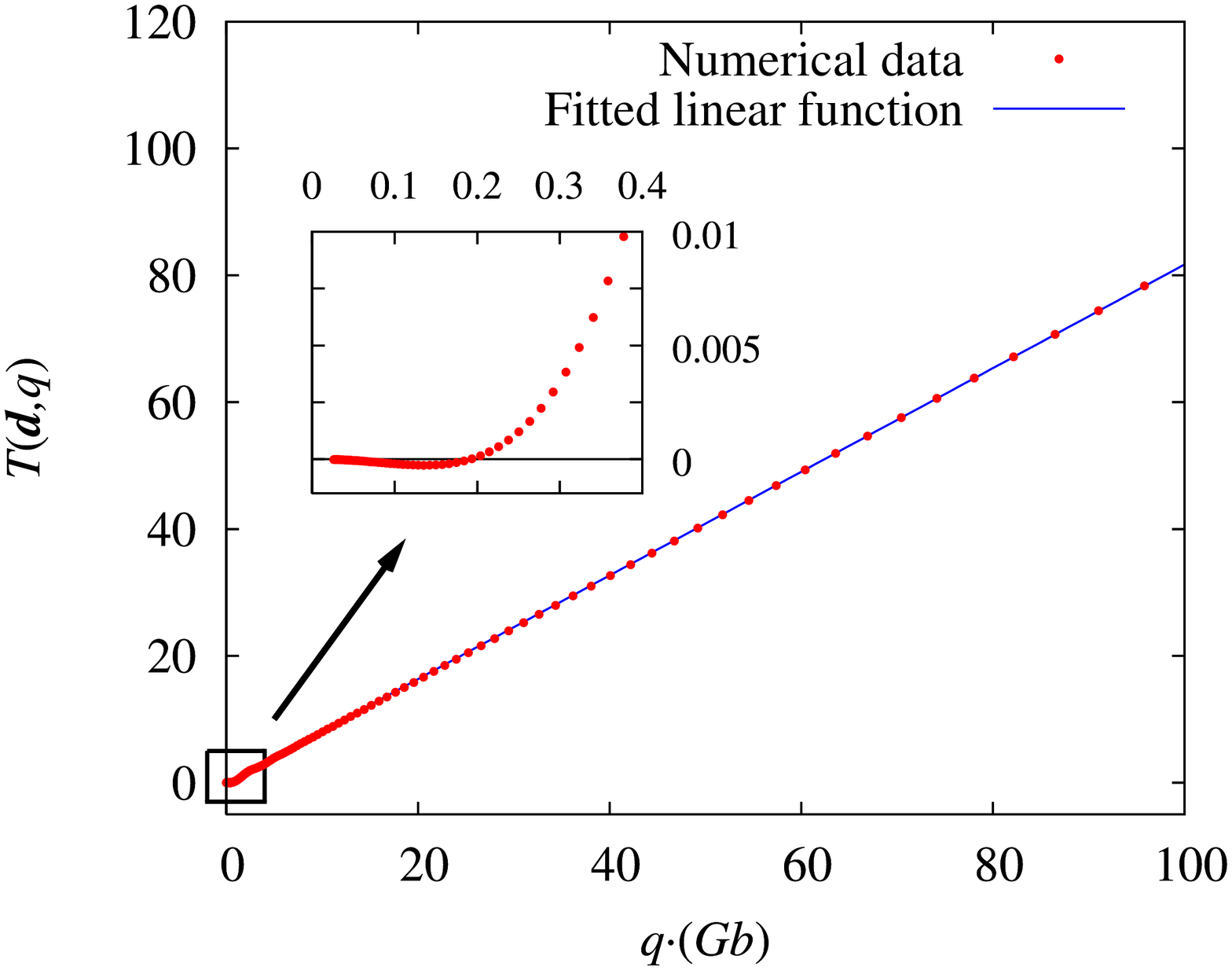}
\caption{\label{fig:T} The $T$ function obtained numerically at $\bm d=\left( \begin{array}{cc} 0.4\\ 0.5 \end{array} \right) $. In the limit of $|q| \rightarrow \infty$ a linear curve was fitted.}
\end{center}
\end{figure}

\begin{figure}[!ht]
\begin{center}
\hspace*{-1.5cm}
\includegraphics[width=6cm,angle=-90]{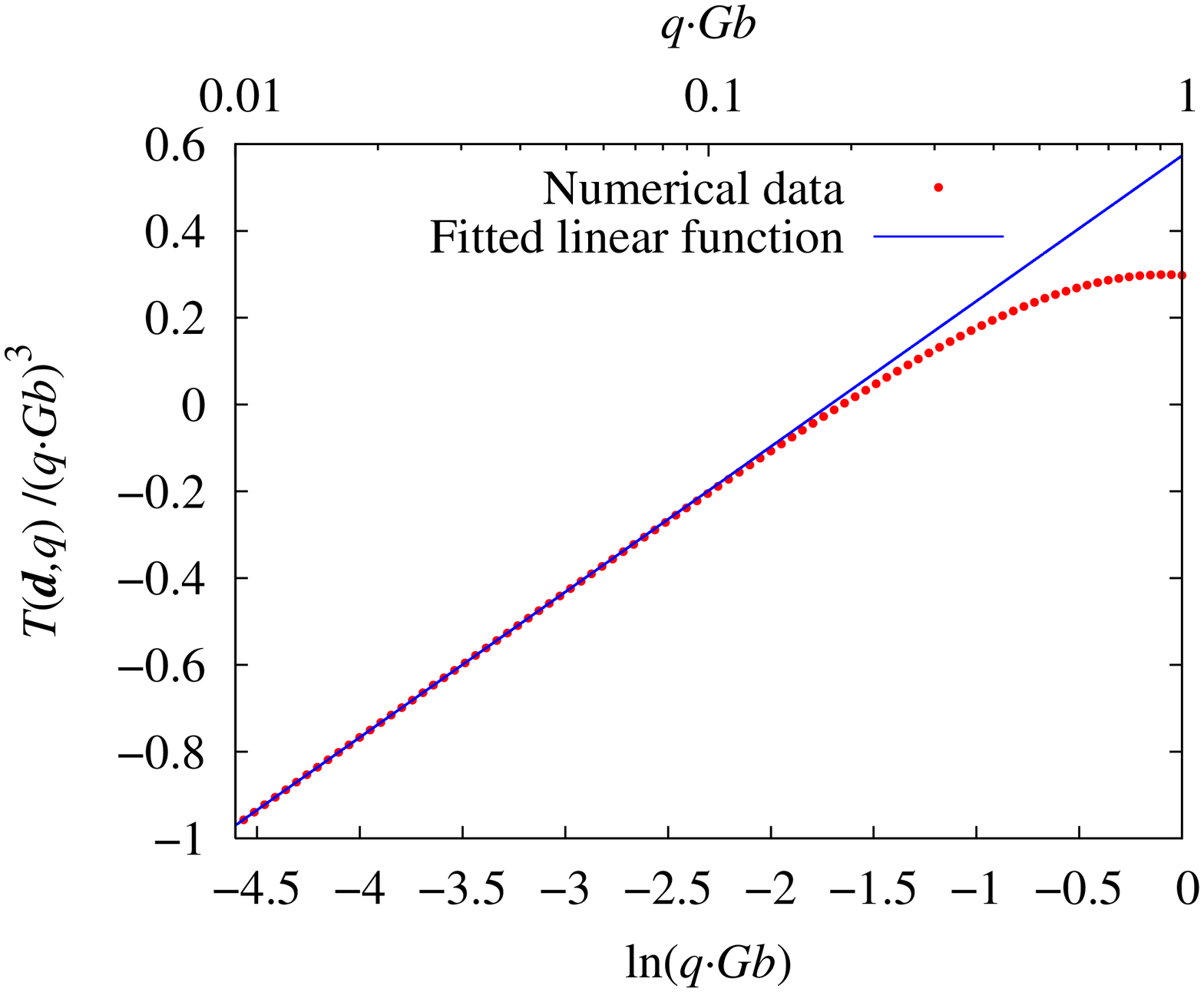}
\caption{\label{fig:T_small_q}$T(\bm d, q)/(Gb\,q^3)$ is plotted as a function of $\ln(q)$ at $\bm d=\left( \begin{array}{cc} 0.4\\ 0.5 \end{array} \right) $. The resulting linear function confirms the $|q| \to 0$ asymptote in (\ref{eqn:T_asymp}).}
\end{center}
\end{figure}

The function $T$ obtained by the method explained above is plotted in figure \ref{fig:T}.
For the two asymptotic regimes $T$ can be well described by the forms:
\begin{eqnarray}
T(\bm d, q)=
\left\{
\begin{array}{ll}
\displaystyle (Gb)^3 \cdot \alpha(\bm d)q^3\ln \left( \frac{q_0(\bm d)}{|q|} \right), & \mathrm{if} \  |q| \to 0,\\
\\
\displaystyle Gb \cdot \beta(\bm d)q, & \mathrm{if} \  |q| \to \infty.
\end{array}
\right.
\label{eqn:T_asymp}
\end{eqnarray}
According to figure \ref{fig:T} the linear relation in (\ref{eqn:T_asymp}) at $|q| \to \infty$ needs no explanation, while the $|q| \to 0$ one is verified in figure \ref{fig:T_small_q}, where $T(\bm d, q)/q^3$ is plotted as a function of $\ln(q)$. The obtained straight line proves (\ref{eqn:T_asymp}). By fitting the above functions to the numerical data of $T$ the values of the parameters $\alpha(\bm d)$, $\beta(\bm d)$ and $q_0(\bm d)$ can be determined for an arbitrary $\bm d$. In table \ref{tab:table1} values for these three parameters at different $\bm d$ values corresponding to dipole angles close to $45^\circ$ are displayed. (The reason for investigating only the vicinity of $45^\circ$ will be discussed in detail in section \ref{sec:stress}.)

\begin{table}
\caption{\label{tab:table1} The values of the $\alpha$, $q_0$ and $\beta$ functions at different $\bm d$-s obtained by fitting to numeric data.}
\begin{indented}
\lineup
\item[]
\begin{tabular}{@{}lllll}
\br
$d_x$ & $d_y$ & \m$\alpha(\bm d)$ & \0$q_0(\bm d)$ & $\beta(\bm d)$\\
\mr
$0.4$ & $0.5$ & $-0.334$ & \0$0.181$ & $0.819$\\
$0.425$ & $0.5$ & $-0.248$ & \0$0.132$ & $0.861$\\
$0.45$ & $0.5$ & $-0.163$ & \0$0.0685$ & $0.902$\\
$0.475$ & $0.5$ & $-0.0807$ & \0$8.90 \cdot 10^{-3}$ & $0.942$\\
$0.5$ & $0.5$ & \0\0 --- $^{\rm a}$ & \0\0--- $^{\rm a}$ & $0.981$\\
$0.525$ & $0.5$ & \m$0.075$ & $46.5$ & $1.021$\\
$0.55$ & $0.5$ & \m$0.147$ & \0$5.55$ & $1.059$\\
$0.575$ & $0.5$ & \m$0.214$ & \0$2.88$ & $1.097$\\
$0.6$ & $0.5$ & \m$0.276$ & \0$2.12$ & $1.135$\\
\br
\end{tabular}
\item[] $^{\rm a}$ The values are not given due to the huge numerical errors.
\end{indented}
\end{table}
The following scaling relation can be easily derived directly from the definition of the $T$ function given by (\ref{eqn:T(d,q)}):
\begin{eqnarray}
T(k \cdot \bm d, q) = k^2 \cdot T\left(\bm d, \frac q{k}\right).
\end{eqnarray}
From this one can find immediately, that according to (\ref{eqn:T_asymp}) the relations
\begin{eqnarray}
\alpha(k \cdot \bm d) &= \frac{\alpha(\bm d)}{k}, \label{eqn:scaling_a}\\
q_0(k \cdot \bm d) &= k \cdot q_0(\bm d), \label{eqn:scaling_q_0}\\
\beta(k \cdot \bm d) &= k \cdot \beta(\bm d) \label{eqn:scaling_b}
\end{eqnarray}
hold for an arbitrary $k \in \mathbb{R}$. Furthermore, as a consequence of (\ref{eqn:T(d,q)})
\begin{eqnarray}
T(\tilde{\bm d}, q) = - T(\bm d, q) \label{eqn:T_tilde}
\end{eqnarray}
is fulfilled too, where $\tilde{\bm d} = \left( \begin{array}{c} -d_x \\ d_y \end{array} \right)$.

According to the relations (\ref{eqn:scaling_a})-(\ref{eqn:T_tilde}) in order to determine the complete form of functions $\alpha$, $\beta$ and $q_0$, one has to investigate their values only at dipole angles between $0$ and $90^\circ$. Using the data from table \ref{tab:table1}, the following approximate functions were found (see figure \ref{fig:T_asymp_params}):

\begin{figure}[!ht]
\begin{center}
\hspace*{-0.5cm}
\includegraphics[width=10cm, angle=-90]{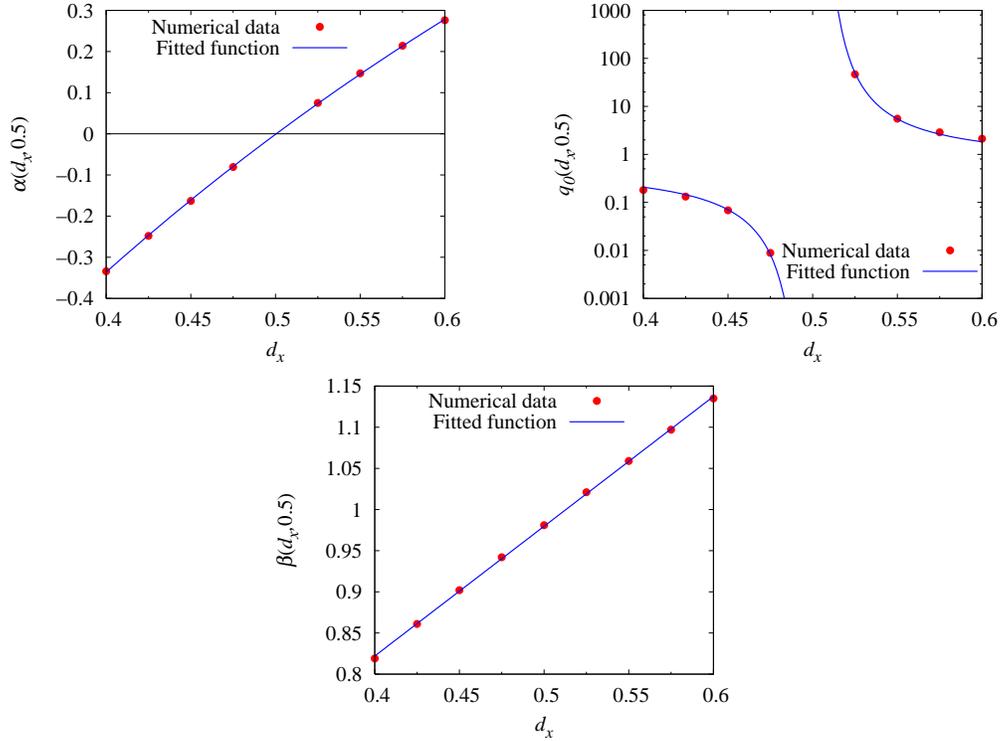}
\end{center}
\caption{\label{fig:T_asymp_params}The numerically obtained values of the functions $\alpha$, $q_0$ and $\beta$, and the fitted functions given by (\ref{eqn:alpha(d_x)})-(\ref{eqn:beta(d_x)}).}
\end{figure}

\begin{eqnarray}
\alpha(d_x, 0.5) = \frac{d_x}{|d_x|} 1.70 \left[ - \exp \bm ( -1.80 ( |d_x| - 0.5) \bm ) + 1.00 \right], \label{eqn:alpha(d_x)}
\end{eqnarray}
\begin{eqnarray}
q_0(d_x, 0.5) = \exp\left( \frac{0.109}{|d_x|-0.500} - 0.486 \right),
\end{eqnarray}
\begin{eqnarray}
\beta(d_x, 0.5) = \frac{d_x}{|d_x|} (1.58 \cdot |d_x| + 0.191).
\label{eqn:beta(d_x)}
\end{eqnarray}

From the relations (\ref{eqn:scaling_a}) and (\ref{eqn:alpha(d_x)}) one gets
\begin{eqnarray}
\alpha(d_x, d_y) &= \alpha\left( \frac{2d_y}{2d_y} d_x, 2d_y \cdot 0.5 \right) = \frac1{2d_y} \alpha\left( \frac{d_x}{2d_y}, 0.5 \right) \nonumber \\
&= \frac{d_x}{|d_x|\cdot |d_y|} 0.85 \left[ -\exp\left( -0.90 \left( \left| \frac{d_x}{d_y} \right| - 1.0 \right) \right) + 1.00 \right].
\label{eqn:a_d}
\end{eqnarray}
The functions $q_0$ and $b$ can be obtained on a similar way leading to
\begin{eqnarray}
q_0(d_x, d_y) = 1.23 \cdot |d_y| \cdot \exp\left( \frac{0.218 \cdot |d_y|}{|d_x|-|d_y|} \right), \label{eqn:q0_d}
\end{eqnarray}
\begin{eqnarray}
\beta(d_x, d_y) = \frac{d_x}{|d_x|} (1.58 \cdot |d_x| + 0.38 \cdot |d_y|). \label{eqn:b_d}
\end{eqnarray}

By combining (\ref{eqn:T_asymp}), (\ref{eqn:a_d}), (\ref{eqn:q0_d}) and (\ref{eqn:b_d}) one concludes
\begin{eqnarray}
\mathrm{Im} \left( \ln(P_{\mathrm{str}}^{F}(q)) \right) =
\left\{
\begin{array}{ll}
\displaystyle (Gb)^3 \frac{\alpha'(\tau_\mathrm{ext})}{2} \rho_\mathrm{dis} q^3\ln \left( \frac{q_0'(\tau_\mathrm{ext})}{|q|} \right), & \mathrm{if} \  |q| \to 0,\\
\\
\displaystyle Gb \frac{\beta'(\tau_\mathrm{ext})}{2} \rho_\mathrm{dis} q, & \mathrm{if} \  |q| \to \infty,
\end{array}
\right.
\label{eqn:im_P2}
\end{eqnarray}
where
\begin{eqnarray}
\alpha'(\tau_\mathrm{ext}) := \frac{\rho_\mathrm{dis}}{2} \int\limits_{\mathbb R^2} d_2^{+-}(\bm d) \alpha(\bm d) \rmd \bm d,
\label{eqn:a_prime}
\end{eqnarray}
\begin{eqnarray}
q_0'(\tau_\mathrm{ext}) := \frac{\int\limits_{\mathbb R^2} d_2^{+-}(\bm d) \alpha(\bm d) \ln(q_0(\bm d)) \rmd \bm d}{\int\limits_{\mathbb R^2} d_2^{+-}(\bm d) \alpha(\bm d) \mathrm d\bm d}
\end{eqnarray}
and
\begin{eqnarray}
\beta'(\tau_\mathrm{ext}) := \frac{\rho_\mathrm{dis}}{2} \int\limits_{\mathbb R^2} d_2^{+-}(\bm d) \beta(\bm d) \rmd \bm d.
\label{eqn:b_prime}
\end{eqnarray}

\section{The effect of the external stress on the microstructure}
\label{sec:stress}
To arrive at the form of the Fourier transform of the stress distribution function that is comparable with the ones obtained on real relaxed dislocation systems the values of the parameters $\alpha'$, $q_0'$ and $\beta'$ have to be determined. According to their definitions given by (\ref{eqn:a_prime})-(\ref{eqn:b_prime}), they depend on the correlation function $d_2^{+-}$. So, in order to get a closed form for the probability distribution $P_{\mathrm{str}}^{F}$, the properties of $d_2^{+-}$ and the microstructure of relaxed dislocation systems have to be analysed.

\subsection{The pair correlation function}
\label{sec:corr}

\begin{figure}[!ht]
\begin{center}
\hspace*{-4cm}(a)\hspace*{7.1cm}(b)

\vspace*{-1.2cm}
\hspace*{-1.5cm}
\includegraphics[width=6.5cm, angle=-90]{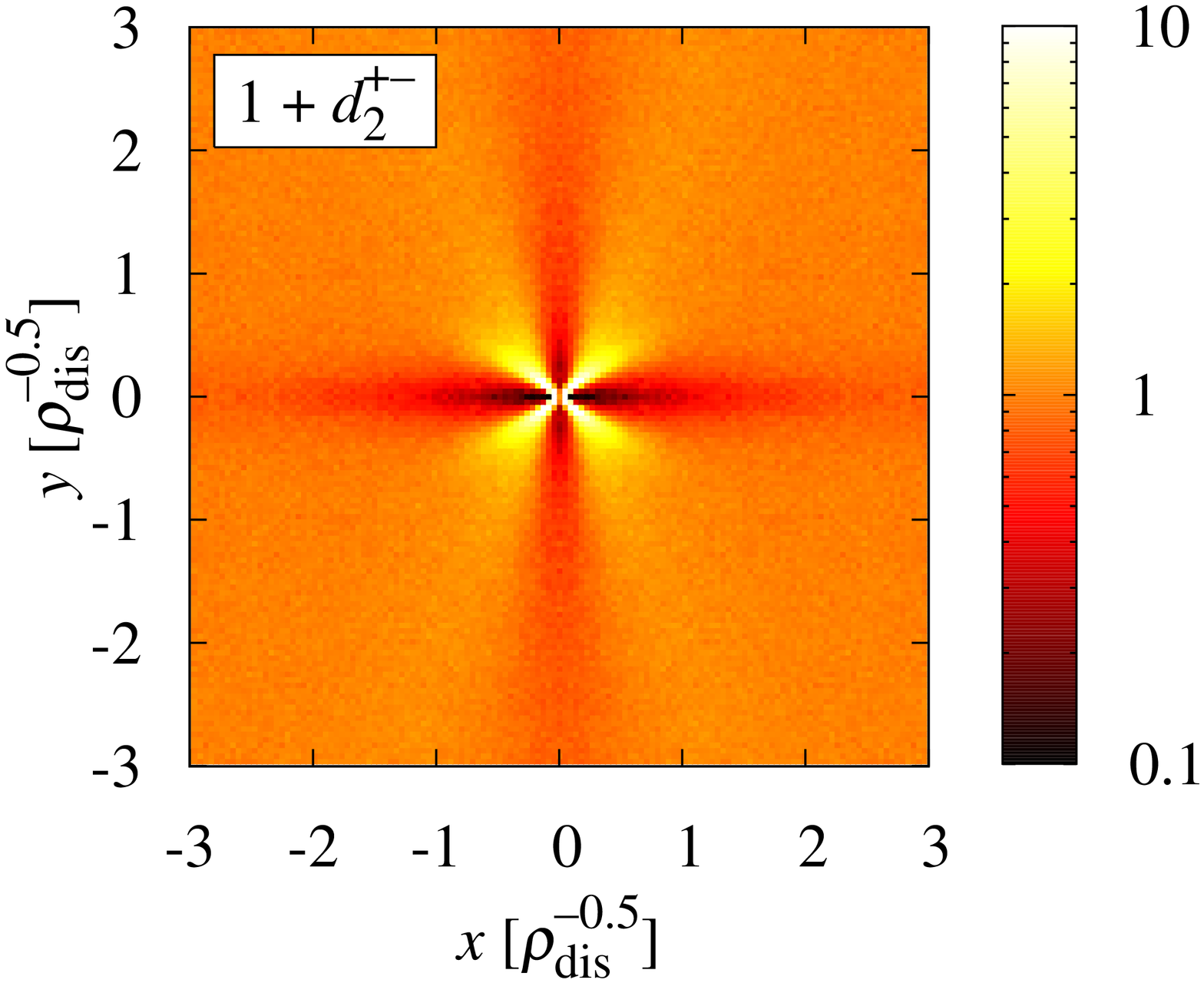}
\hspace*{-2cm}
\includegraphics[width=6.5cm, angle=-90]{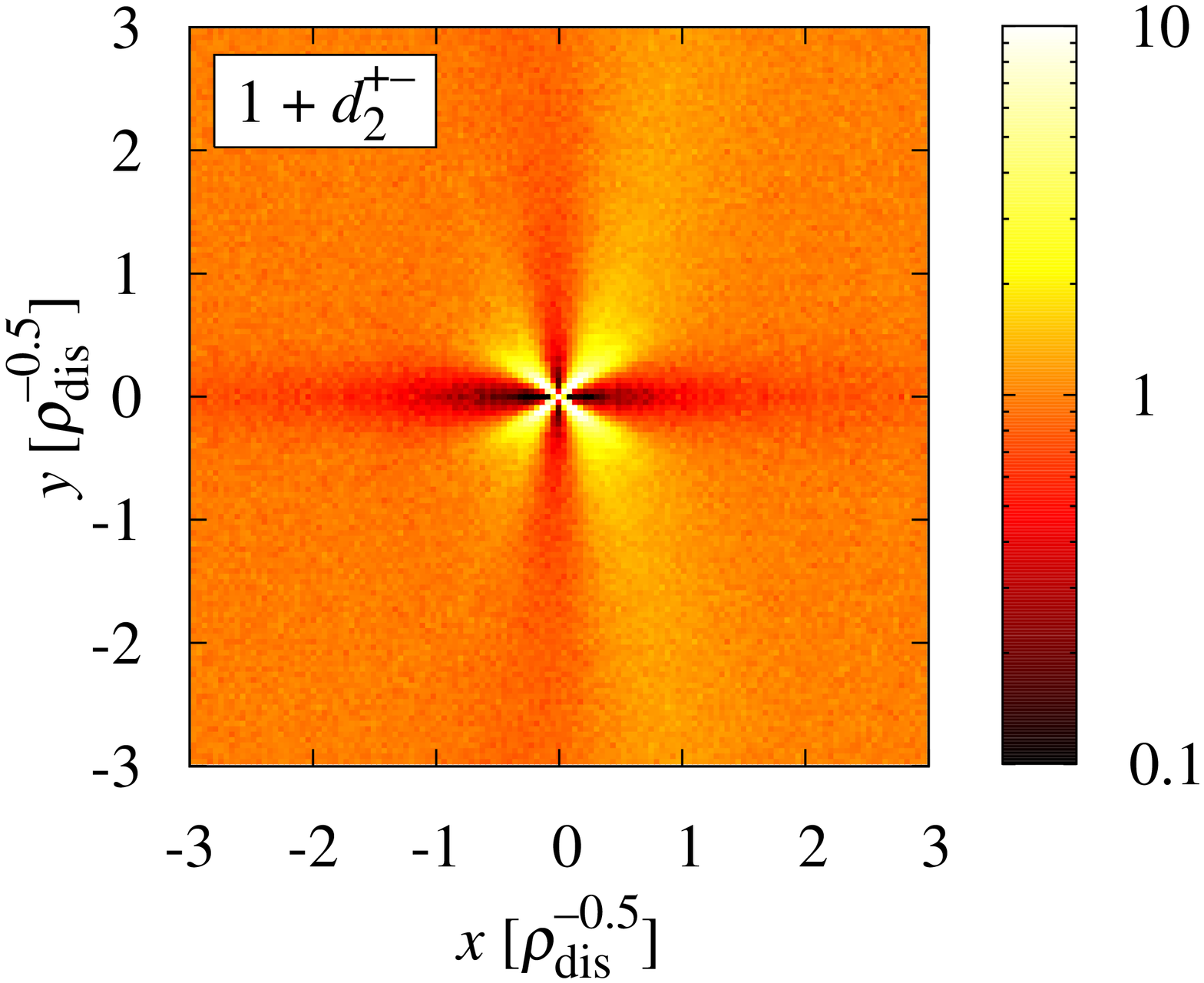}
\hspace*{-1cm}

\hspace*{-3.5cm}(c)

\vspace*{-1.2cm}
\includegraphics[width=6.5cm, angle=-90]{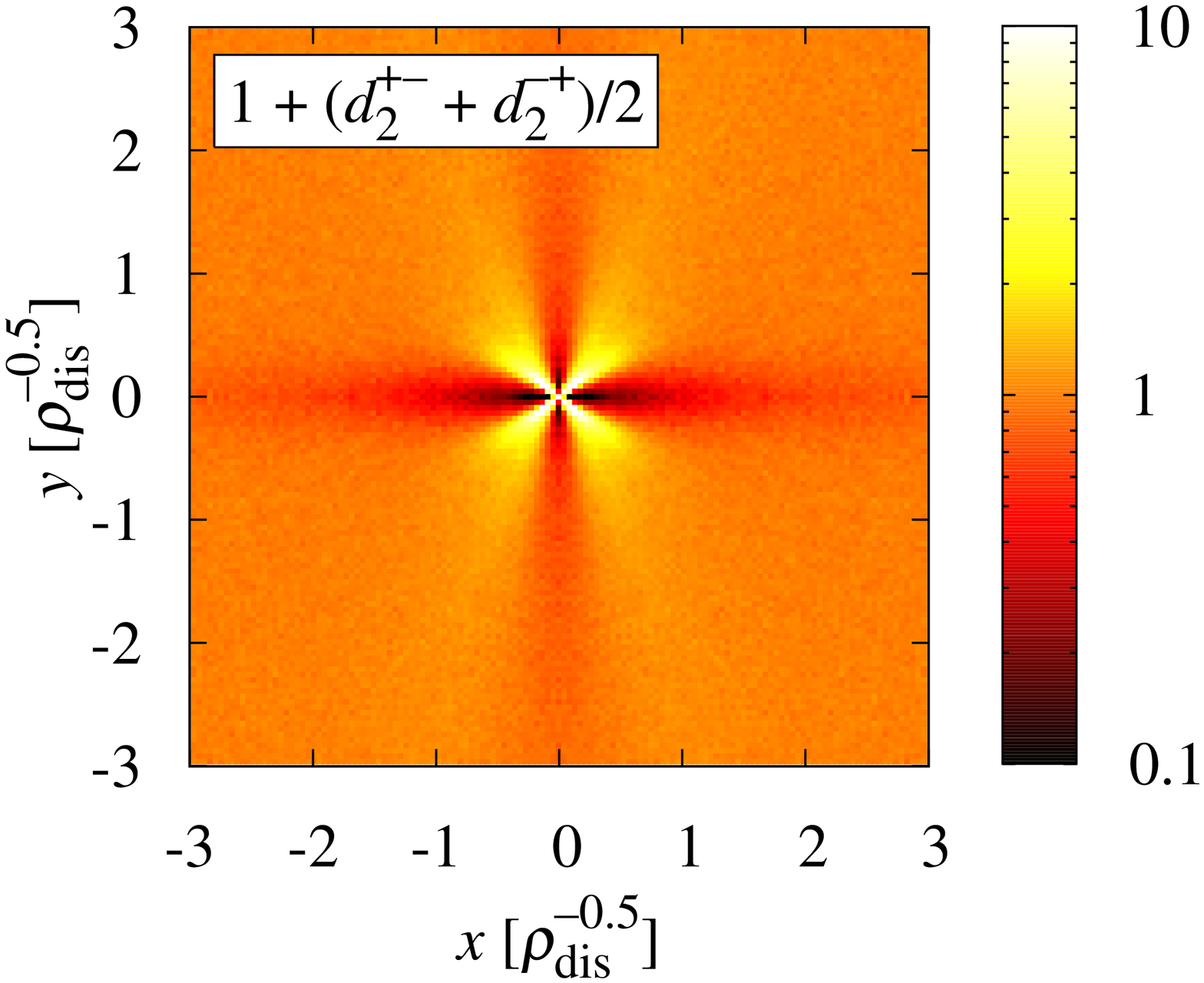}
\end{center}
\caption{\label{fig:corr} The $1+d_2^{+-}$ correlation functions obtained numerically from many different relaxed dislocation configurations, (a) at zero external stress (b) at $\tau_\mathrm{ext} = (1/\sqrt{128}) \cdot G b \sqrt{\rho_\mathrm{dis}} \approx 0.09 \cdot G b \sqrt{\rho_\mathrm{dis}}$ external shear stress. (c) $1+(d_2^{+-}+d_2^{-+})/2$ is also plotted at $\tau_\mathrm{ext} = (1/\sqrt{128}) \cdot G b \sqrt{\rho_\mathrm{dis}} \approx 0.09 \cdot G b \sqrt{\rho_\mathrm{dis}}$ to demonstrate, that it practically does not change due to the external stress. In all cases averaging was performed over 2000 different realizations.}
\end{figure}

The $d_2^{+-}$ correlation function of a relaxed homogeneous dislocation system can be directly determined by discrete dislocation dynamics simulations by counting the relative coordinates between positive and negative dislocations in the equilibrium configuration obtained numerically. In order to get smooth correlation function one has to perform averaging over many different realizations. Figure \ref{fig:corr}(a) shows the correlation function at zero external stress. In figure \ref{fig:corr}(b) $d_2^{+-}$ at external shear stress $\tau_\mathrm{ext} = (1/\sqrt{128}) \cdot Gb \sqrt{\rho_\mathrm{dis}} \approx 0.09 \cdot Gb \sqrt{\rho_\mathrm{dis}}$ is plotted. In both cases the simulations were started from a random distribution of 64 positive and 64 negative dislocations and 2000 different realizations were used for averaging.

One can see in figure \ref{fig:corr} that in relaxed systems the positive and negative dislocations tend to form narrow dislocation dipoles. Applied external stress makes $d_2^{+-}$ asymmetric but its nature close to the origin remains unchanged. The reason for this is that splitting a narrow dipole requires very high stress. Even when the acting stress is a bit below the yield stress, only wide dipoles are split up (in the 2D single slip dislocation systems under study the yield stress is $\tau_\mathrm{y} \approx 0.1 \cdot G b \sqrt{\rho_\mathrm{dis}}$ \cite{miguel3}). We note that the symmetric part of $d_2^{+-}$ [which is equal to $(d_2^{+-}+d_2^{-+})/2$] is nearly unaffected by the external stress (see figure \ref{fig:corr}(c)).

\begin{figure}[!ht]
\begin{center}
\hspace*{-4cm}(a)\hspace*{7.1cm}(b)

\vspace*{-1.2cm}
\hspace*{-1.5cm}
\includegraphics[width=6.5cm, angle=-90]{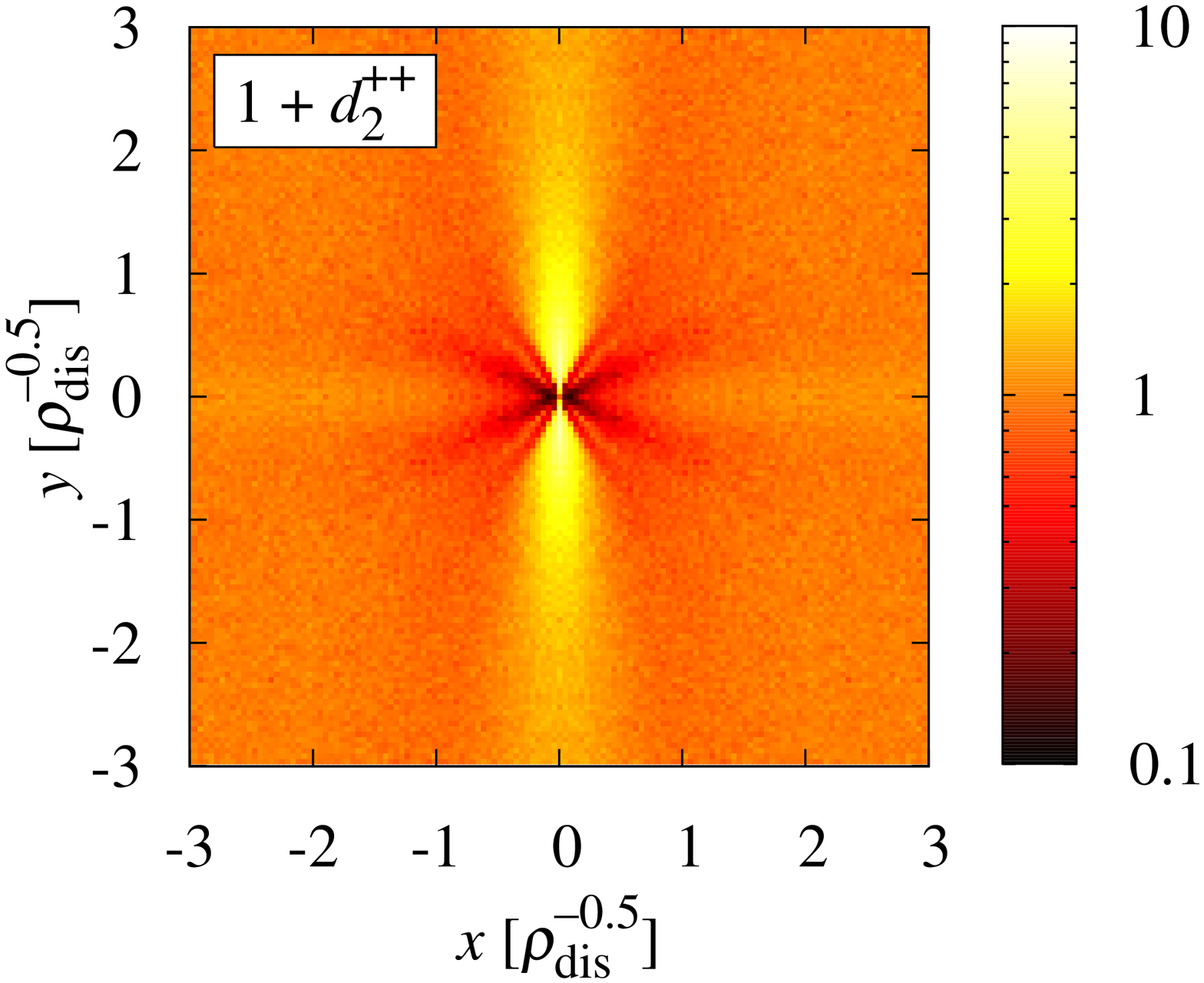}
\hspace*{-2cm}
\includegraphics[width=6.5cm, angle=-90]{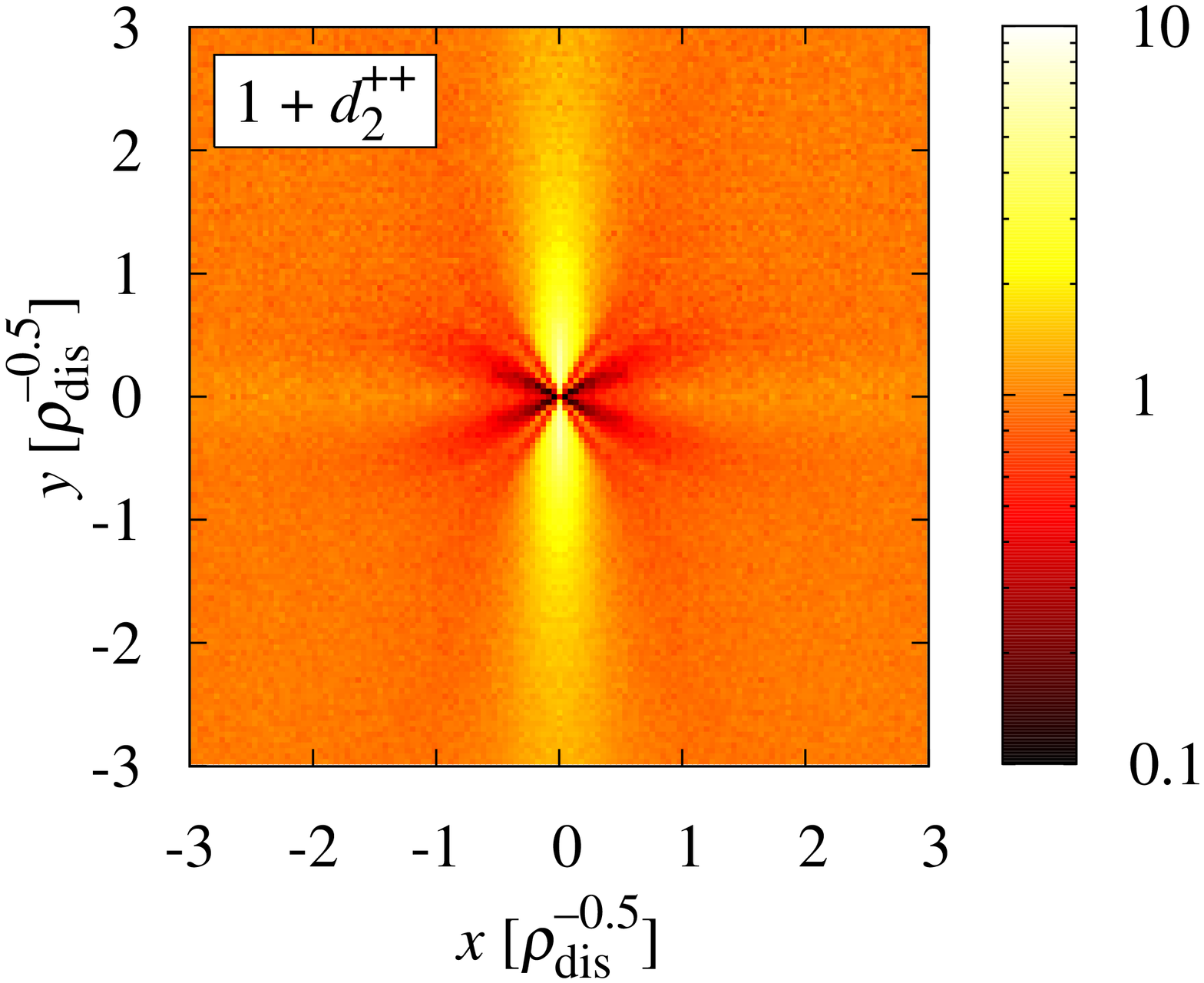}
\hspace*{-1cm}
\end{center}
\caption{\label{fig:corr_pp} The $1+d_2^{++}$ correlation functions obtained numerically from 2000 different relaxed dislocation configurations. Plot (a) and (b) correspond to the same stress level as the two in figure \ref{fig:corr}.}
\end{figure}

\begin{figure}[!ht]
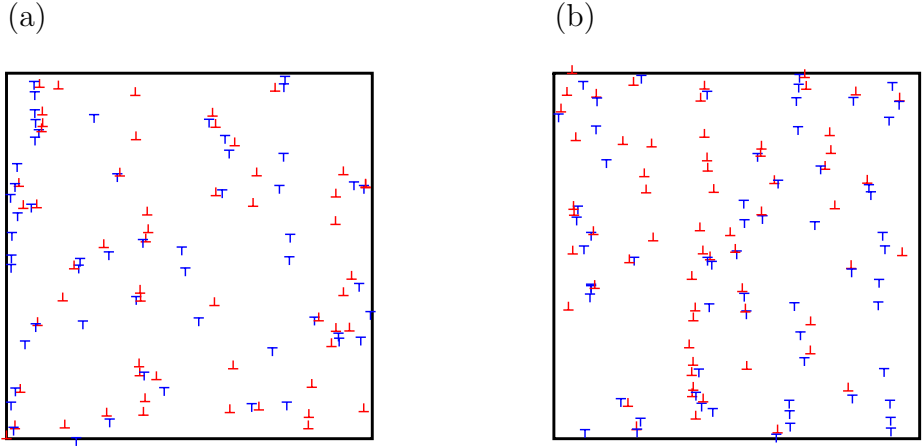

\begin{center}
\hspace*{-4.3cm}(a)\hspace*{6.75cm}(b)

\includegraphics[width=5cm, angle=-90]{fig7a.ps}
\hspace*{2cm}
\includegraphics[width=5cm, angle=-90]{fig7b.ps}
\end{center}
\caption{\label{fig:relaxed} Relaxed dislocation configurations obtained by discrete dislocation simulations. In figure (a) there was no applied shear stress, while in figure (b) the external shear stress was $\tau_\mathrm{ext} = (1/\sqrt{128}) \cdot G b \sqrt{\rho_\mathrm{dis}} \approx 0.09 \cdot G b \sqrt{\rho_\mathrm{dis}}$. Notice that in the stressed system the narrow dipoles are still present as well as the dislocation walls.}
\end{figure}

The $d_2^{++}$ correlation function was also plotted in figure \ref{fig:corr_pp} to demonstrate that it practically does not change due to external stress, and that dislocation walls are also present in stressed configurations. In figure \ref{fig:relaxed} a stressed and an unstressed relaxed configuration is shown. The presence of the mentioned dislocation dipoles and walls can be clearly seen.

\subsection{Monodisperse dislocation dipole systems}
\label{sec:monod}

The properties of the $d_2^{+-}$ correlation function mentioned above and the relaxed configurations plotted in figure \ref{fig:relaxed} indicates that mainly dislocation dipoles affect the values of $\alpha'$, $q_0'$ and $\beta'$, both in stressed and unstressed systems [see equations (\ref{eqn:a_prime})-(\ref{eqn:b_prime})]. So, in this section as a first step we consider a set of randomly positioned ideal $45^\circ$ dipoles with same momenta. This will be referred to as a monodisperse dislocation dipole system. It is important to note, that this is only a model system, since it is not in a dynamic equilibrium. As a result of random positions, the dipoles may even overlap. Although, this approach may seem to be oversimplified at first glance, as it is demonstrated later the results obtained can capture the properties of the stress distribution function found on more realistic configurations. Similar method was used earlier by Csikor and Groma \cite{csikor}. In their paper the relaxed 2D distribution was envisaged as a set of ideal dislocation dipoles and short ideal walls. Using this concept the form of the distribution function was given analytically for the stress free case. The resulting theoretical distribution function was in complete agreement with the one obtained numerically. Since small external stress does not split up the dipoles, it is reasonable to adopt this approach for the loaded case considered in this paper.

\begin{figure}[!ht]
\psfrag{####ep###}{\hspace*{-0.3cm}$\varepsilon^+(\tau_\mathrm{ext}, y_0)$}
\psfrag{####em###}{\hspace*{-0.3cm}$\varepsilon^-(\tau_\mathrm{ext}, y_0)$}
\psfrag{dy}{$y_0$}
\psfrag{##dx##}{$y_0$}
\psfrag{##d##}{$\bm d$}
\psfrag{##d\'##}{$\bm d'$}
\psfrag{#a}{(a)}
\psfrag{#b}{(b)}
\psfrag{#c}{(c)}
\psfrag{#d}{(d)}
\begin{center}
\includegraphics[width=11cm]{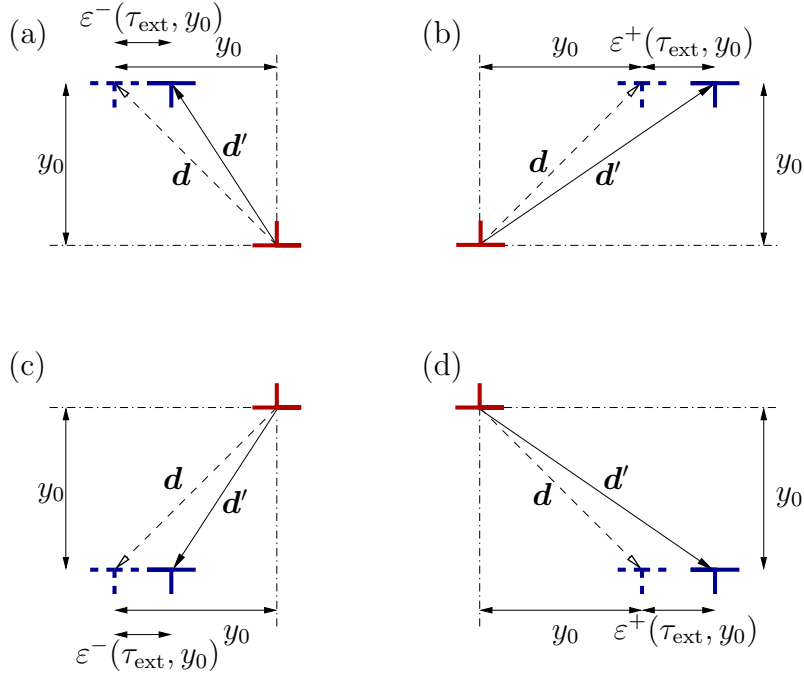}
\caption{\label{fig:dipole_def} Under external shear stress $\tau_\mathrm{ext}$ the stress free dipole moment $\bm d$ becomes $\bm d'$. The change of the $x$ coordinate is denoted by $\varepsilon^-(\tau_\mathrm{ext}, y_0)$ for (a) $135^\circ$ and (c) $225^\circ$ dipoles and by $\varepsilon^+(\tau_\mathrm{ext}, y_0)$ for (b) $45^\circ$ and (d) $315^\circ$ dipoles, where $y_0 = d_y$.}
\end{center}
\end{figure}

We start our analysis by studying the behaviour of an isolated dipole subjected to shear stress. In equilibrium the sum of the forces (the Peach-Koehler force and the force produced by the external stress) acting on both dislocations must be zero. If there is no external stress there are four possible stable dislocation dipole orientations corresponding to the dipole angles $45^\circ$, $135^\circ$, $225^\circ$ and $315^\circ$. If external stress is applied let $\varepsilon^\pm(\tau_\mathrm{ext}, y_0)$ denote the change of the $x$ coordinate of the negative dislocation (see figure \ref{fig:dipole_def}) with $y_0$ denoting the distance of the slip planes of the dislocations.

\begin{figure}[!ht]
\begin{center}
\hspace*{-1.5cm}
\includegraphics[width=6cm, angle=-90]{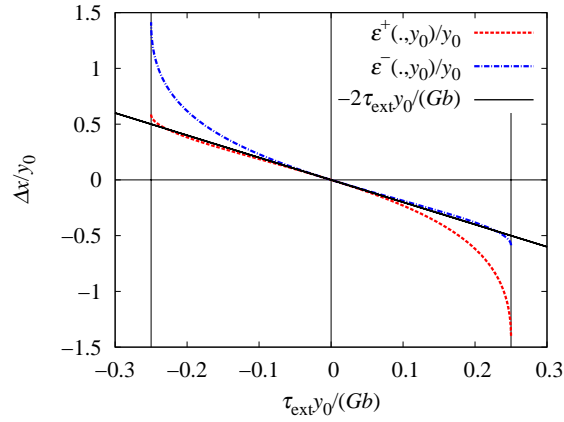}
\caption{\label{fig:epsilon} The $\varepsilon^\pm$ functions. Their tangent in the origin given by (\ref{eqn:epsilon_origin}) is also plotted.}
\end{center}
\end{figure}

The $\varepsilon^\pm(\tau_\mathrm{ext}, y_0)$ functions can be determined from (\ref{eqn:tauind}) (see figure \ref{fig:epsilon}). It is easy to see that
\begin{eqnarray}
\varepsilon^\pm(\tau_\mathrm{ext}, y_0) \approx -2 \frac{\tau_\mathrm{ext} y_0^2}{Gb} \mathrm{\quad if \ } |\tau_\mathrm{ext}| \to 0.
\label{eqn:epsilon_origin}
\end{eqnarray}
It should be noted that the domain of $\varepsilon^\pm(.,y_0)$ is the interval $\left[-\frac{Gb}{4y_0},\frac{Gb}{4y_0} \right]$ since at higher stresses the dipole splits.

The monodisperse dipole system under external stress is built up of randomly positioned, separately deformed dipoles, as in figure \ref{fig:epsilon}. The main advantage of investigating a monodisperse system is that the corresponding $d_2^{+-}$ correlation function can be exactly given. If there are $N/2$ positive and $N/2$ negative dislocations then
\begin{eqnarray}
d_2^{+-}(\bm r) = \frac{2}{\rho_{\mathrm{dis}}} \delta(y-y_0) \delta \bm ( x - y - \varepsilon^+(\tau_\mathrm{ext}, y_0) \bm ) - \frac{2}{N},
\label{eqn:d_monod}
\end{eqnarray}
where $y_0 > 0$.

From (\ref{eqn:a_d}), (\ref{eqn:d_monod}) and (\ref{eqn:a_prime}) we arrive at
\begin{eqnarray}
\alpha'(\tau_\mathrm{ext}) &= \frac{\rho_\mathrm{dis}}{2} \int\limits_{\mathbb R^2} d_2^{+-}(\bm r) \alpha(\bm r) \rmd \bm r = \int\limits_{\mathbb R^2} \delta(y-y_0) \delta(x - y - \varepsilon^+(\tau_\mathrm{ext}, y_0)) \nonumber \\
&\quad \times \frac{x}{|x|\cdot |y|} 0.85 \left[ -\exp\left( -0.90 \left( \left| \frac{x}{y} \right| - 1.0 \right) \right) + 1.00 \right] \rmd x \, \rmd y \nonumber \\
&= 0.85 \int\limits_{\mathbb R} \delta(y-y_0) \frac{1}{|y|} \left[ -\exp\left( \frac{-0.90}{|y|} \varepsilon^+(\tau_\mathrm{ext}, y_0) \right) + 1.00 \right] \rmd y \nonumber \\
&= \frac{0.85}{|y_0|} \left[ -\exp\left( \frac{-0.90}{|y_0|} \varepsilon^+(\tau_\mathrm{ext}, y_0) \right) + 1.00 \right] \nonumber \\
&\approx \frac{0.85\cdot 0.90}{|y_0|^2} \varepsilon^+(\tau_\mathrm{ext}, y_0) \approx -1.53 \frac{\tau_\mathrm{ext}}{Gb}
\end{eqnarray}
\begin{eqnarray}
\beta'(\tau_\mathrm{ext}) &= \frac{\rho_\mathrm{dis}}{2} \int\limits_{\mathbb R^2} d_2^{+-}(\bm r) \beta(\bm r) \rmd \bm r = \int\limits_{\mathbb R^2} \delta(y-y_0) \delta(x - y - \varepsilon^+(\tau_\mathrm{ext}, y_0)) \nonumber \\
&\quad \times \frac{x}{|x|} \left( 1.58 \cdot |x| + 0.38\cdot |y| \right) \rmd x \, \rmd y \nonumber \\
&= 1.58 \cdot [y_0 + \varepsilon^+(\tau_\mathrm{ext}, y_0) ] + 0.38 \cdot y_0 \approx 1.96\cdot y_0 - 3.16 \frac{\tau_\mathrm{ext}}{Gb} y_0^2.
\end{eqnarray}
The above results correspond to configurations consisting of only $45^\circ$ dipoles. After a short and straightforward calculation one finds that for systems consisting of the four possible ideal dipoles in equal number:
\begin{eqnarray}
\alpha'(\tau_\mathrm{ext}) \approx -1.53 \frac{\tau_\mathrm{ext}}{Gb}, \label{eqn:a_prime_monod} \\
\beta'(\tau_\mathrm{ext}) \approx -3.16 \frac{\tau_\mathrm{ext}}{Gb} y_0^2. \label{eqn:b_prime_monod}
\end{eqnarray}
(The $q_0'$ function could be determined in a similar way but it will not play any role in the further investigations so, its actual form is not given explicitly.)

\subsection{Disperse dislocation dipole systems}

A more realistic model of a 2D relaxed dislocation systems is when we assume that the dipole momenta of the dislocation dipoles varies according to a given distribution. Let $P_\mathrm{dip}(y_0)$ denote the probability distribution of the dipole height $y_0$. Like in the monodisperse systems studied above (where $P_\mathrm{dip}(y_0)$ is simply a Dirac delta function), the $d_2^{+-}$ correlation function can be explicitly given by
\begin{eqnarray}
d_2^{+-}(\bm r) = \frac{2}{\rho_{\mathrm{dis}}} P_\mathrm{dip}(y) \delta(x - y - \varepsilon^+(\tau_\mathrm{ext}, y))-\frac{2}{N}.
\label{eqn:d_disp}
\end{eqnarray}
For the case of the four different types of dipoles the correlation function could be given accordingly. From (\ref{eqn:a_prime_monod}) and (\ref{eqn:b_prime_monod}) one concludes that
\begin{eqnarray}
\alpha'(\tau_\mathrm{ext}) \approx -1.53 \frac{\tau_\mathrm{ext}}{Gb}, \\
\beta'(\tau_\mathrm{ext}) \approx -3.16 \frac{\tau_\mathrm{ext}}{Gb} \int\limits_{\mathbb R} P_\mathrm{dip}(y) y^2 \rmd y.
\end{eqnarray}

\subsection{Model of real relaxed dislocation configurations}
\label{sec:model_relaxed}

As it was mentioned above, relaxed 2D configurations are made up of two basic structures, dipoles and walls. In a simple picture, a certain fraction of the dislocations forms dipoles with different momenta. Let us denote this ratio by $K$. The remaining dislocations are distributed in the walls. This means that the correlation function given by (\ref{eqn:d_disp}) of the disperse case has to be simply multiplied by $K$:
\begin{eqnarray}
d_2^{+-}(\bm r) = \frac{2K}{\rho_{\mathrm{dis}}} P_\mathrm{dip}(y) \delta(x - y - \varepsilon^+(\tau_\mathrm{ext}, y))-\frac{2K}{N}
\label{eqn:d_relax}
\end{eqnarray}
and so for systems containing all four ideal dipole types in equal number
\begin{eqnarray}
\alpha'(\tau_\mathrm{ext}) \approx -1.53 \frac{K \tau_\mathrm{ext}}{Gb}, \label{eqn:a_prime_real} \\
\beta'(\tau_\mathrm{ext}) \approx -3.16 \frac{K \tau_\mathrm{ext}}{Gb} \int\limits_{\mathbb R} P_\mathrm{dip}(y) y^2 \rmd y. \label{eqn:b_prime_real}
\end{eqnarray}

Csikor and Groma determined the value of $K$ by investigating relaxed configurations \cite{csikor}. They looked at nearest neighbour dislocations, and measured how frequently their signs are opposite. According to their result
\begin{eqnarray}
K = 0.76 \pm 0.01\,.
\end{eqnarray}
A $P_\mathrm{dip}(y)$ dipole height distribution was also established by measuring the distance of the nearest dislocation of opposite sign for each dislocations. For relaxed configurations
\begin{eqnarray}
P_\mathrm{dip}(y) = \frac 1{2 y_0} \exp\left( -\frac{|y|}{y_0} \right) \qquad \mathrm{with } \qquad y_0 \approx \frac{0.35}{\sqrt{\rho_{\mathrm{dis}}}}
\end{eqnarray}
was found. With these, from (\ref{eqn:a_prime_real}) and (\ref{eqn:b_prime_real})
\begin{eqnarray}
\alpha'(\tau_\mathrm{ext}) \approx -1.16 \cdot \frac{\tau_\mathrm{ext}}{Gb},
\label{eqn:a_prime_real2}
\end{eqnarray}
and
\begin{eqnarray}
\beta'(\tau_\mathrm{ext}) \approx -\frac{0.60}{\rho_\mathrm{dis}} \frac{\tau_\mathrm{ext}}{Gb}.
\label{eqn:b_prime_real2}
\end{eqnarray}

\section{The evolving distribution function}
\label{sec:evolv}

In the considerations given above the asymptotes of the Fourier transform of the distribution function
are given. For its real part the result obtained by Csikor and Groma given by (\ref{eqn:re_P}) \cite{csikor} holds for the loaded case too, while for the complex part (\ref{eqn:im_P2}) was derived. By combining (\ref{eqn:re_P}) and (\ref{eqn:im_P2}) we arrive at
\begin{eqnarray}
\fl P_{\mathrm{str}}^{F}(q) = \left\{
\begin{array}{ll}
\displaystyle 1 + C \rho_{\mathrm{dis}} q^2 \ln\left( \frac{|q|}{q_\mathrm{eff}} \right) + \rmi \, (Gb)^3 \frac{\alpha'(\tau_\mathrm{ext})}{2} \rho_\mathrm{dis} q^3\ln \left( \frac{q_0'(\tau_\mathrm{ext})}{|q|} \right), & \mathrm{if} \  |q| \to 0,\\
\\
\displaystyle \exp \left( -\frac{D}{2} \rho_\mathrm{dis} |q| \right) \exp \left( \rmi \, Gb \frac{\beta'(\tau_\mathrm{ext})}{2} \rho_{\mathrm{dis}} q \right), & \mathrm{if} \  |q| \to \infty.
\end{array}
\right.
\label{eqn:P_F_tot}
\end{eqnarray}
The $\alpha'$ and $\beta'$ functions, appearing in the above expressions, were calculated in section \ref{sec:stress} for different model dislocation systems. (The $q_0'$ function does not have any effect on the asymptotes of the resulting distribution function, so it was not given explicitly.)

As the last step, in this section the asymptotic decay of $P_\mathrm{str}(\tau)$ and the behaviour of its centre part is determined from its Fourier transform given by (\ref{eqn:P_F_tot}).

\subsection{Behaviour of the third order restricted moment and the asymptotic decay of the distribution function}

Let us divide the $|q| \to 0$ case expression in (\ref{eqn:P_F_tot}) into two terms:
\begin{eqnarray}
P_{\mathrm{str}}^{F}(q) = P_{\mathrm{str},1}^{F}(q) + P_{\mathrm{str},2}^{F}(q) \label{eqn:p_div}
\end{eqnarray}
with
\begin{eqnarray}
P_{\mathrm{str},1}^{F}(q) := 1 + C \rho_{\mathrm{dis}} q^2 \ln\left( \frac{|q|}{q_\mathrm{eff}} \right)
\label{eqn:p_div2}
\end{eqnarray}
and
\begin{eqnarray}
P_{\mathrm{str},2}^{F}(q) := \rmi \, (Gb)^3 \frac{\alpha'(\tau_\mathrm{ext})}{2} \rho_\mathrm{dis} q^3\ln \left( \frac{q_0'(\tau_\mathrm{ext})}{|q|} \right).
\end{eqnarray}
By analysing (\ref{eqn:p_div2}) Groma and Bak\'o \cite{gromabako} have found that
\begin{eqnarray}
P_{\mathrm{str},1}(\tau) = C \rho_{\mathrm{dis}} \frac{1}{|\tau|^3}~,\quad\mathrm{if} \  |\tau| \to \infty.\label{eqn:tau3}
\end{eqnarray}

In order to find $P_{\mathrm{str},2}$, first, we recall the well known fact that for every distribution function $p$ differentiable $k$ times the identity
\begin{eqnarray}
m_k = \rmi^k (p^F)^{(k)}(0) \label{eqn:moment}
\end{eqnarray}
holds, where $m_k$ is the $k$th order moment of $p$:
\begin{eqnarray}
m_k := \int\limits_{-\infty}^\infty x^k p(x) \rmd x,
\end{eqnarray}
and $(\cdot)^{(k)}$ denotes the $k$th derivative.

We have to note, that the above relation cannot be applied for $P_{\mathrm{str},2}^{F}$ since its 3rd derivative is infinite at $q=0$. It can be seen, however, if we define the function
\begin{eqnarray}
\widetilde{P_{\mathrm{str},2}^{F}}(q) := P_{\mathrm{str},2}^{F}(q) - \frac{1}{2^3} P_{\mathrm{str},2}^{F}(2q) = \rmi \, (Gb)^3 \frac{\alpha'(\tau_\mathrm{ext})}{2} \rho_\mathrm{dis} q^3 \ln 2,
\end{eqnarray}
unlike $P_{\mathrm{str},2}^{F}$, it has a finite 3rd derivative at zero. So, the relation (\ref{eqn:moment}) can be applied with $k=3$ leading to 
\begin{eqnarray}
\lim_{\tau \to \infty} \int\limits_{-\tau}^{\tau} (\tau')^3 \widetilde{P_{\mathrm{str},2}}(\tau') \mathrm d \tau' = 3 (Gb)^3 \alpha'(\tau_\mathrm{ext}) \rho_{\mathrm{dis}} \ln 2.
\label{eqn:moment1}
\end{eqnarray}
On the other hand
\begin{eqnarray}
\int\limits_{-\tau}^{\tau} (\tau')^3 \widetilde{P_{\mathrm{str},2}}(\tau') \rmd \tau' = \int\limits_{-\tau}^{\tau} (\tau')^3 P_{\mathrm{str},2}(\tau') \rmd \tau' - \frac{1}{2^4} \int\limits_{-\tau}^{\tau} (\tau')^3 P_{\mathrm{str},2}(\tau'/2) \rmd \tau' \nonumber \\
\quad = v_3(\tau)-v_3(\tau/2),
\label{eqn:moment2}
\end{eqnarray}
where the function
\begin{eqnarray}
v_3(\tau) := \int\limits_{-\tau}^{\tau} (\tau')^3 P_{\mathrm{str},2}(\tau') \rmd \tau',
\label{eqn:v3}
\end{eqnarray}
called the third order restricted moment, is introduced. After combining (\ref{eqn:moment1}) and (\ref{eqn:moment2}) one gets:
\begin{eqnarray}
\lim_{\tau \to \infty} \left[ v_3(\tau)-v_3(\tau/2) \right] = 3 (Gb)^3 \alpha'(\tau_\mathrm{ext}) \rho_{\mathrm{dis}} \ln 2,
\end{eqnarray}
which is a function equation for $v_3(\tau)$. Its general solution is
\begin{eqnarray}
\lim_{\tau \to \infty} v_3(\tau)=3 (Gb)^3 \alpha'(\tau_\mathrm{ext}) \rho_{\mathrm{dis}} \ln\left( \frac{\tau}{\tau_0} \right),
\label{eqn:v3_tail}
\end{eqnarray}
where $\tau_0$ is an arbitrary constant. It follows that
\begin{eqnarray}
P_{\mathrm{str},2}(\tau) = \frac{3 (Gb)^3 \alpha'(\tau_\mathrm{ext}) \rho_{\mathrm{dis}}}{2 \tau|\tau|^3}, \quad\mathrm{if} \  |\tau| \to \infty.
\label{eqn:P_str,2}
\end{eqnarray}

Finally, from (\ref{eqn:p_div}), (\ref{eqn:tau3}) and (\ref{eqn:P_str,2}) one concludes that
\begin{eqnarray}
P_{\mathrm{str}}(\tau) = P_{\mathrm{str},1}(\tau) + P_{\mathrm{str},2}(\tau) = \frac{C \rho_{\mathrm{dis}}}{|\tau|^3} + \frac{U \rho_{\mathrm{dis}}}{\tau|\tau|^3}, \quad\mathrm{if} \  |\tau| \to \infty,
\label{eqn:P_asymp_final}
\end{eqnarray}
where
\begin{eqnarray}
U = \frac 32 (Gb)^3 \alpha'(\tau_\mathrm{ext}).
\label{eqn:U_def}
\end{eqnarray}

\subsection{The central part of the distribution function}
\label{sec:central}

According to (\ref{eqn:P_F_tot}), in the limit of $|q| \to \infty$, due to the external stress, the Fourier transform is multiplied by the phase factor $\exp \left( \rmi \, Gb \frac{\beta'(\tau_\mathrm{ext})}{2} \rho_{\mathrm{dis}} q \right)$. From the identity
\begin{eqnarray}
\frac{1}{2\pi} \int\limits_\mathbb{R} f(x+x_0) \rme^{-\rmi q x} \rmd x = f^F(q) \rme^{\rmi q x_0}
\label{eqn:phase}
\end{eqnarray}
valid for an arbitrary function $f$, one can speculate that at $|\tau|\to 0$ the distribution function is shifted with $\Delta \tau := Gb \frac{\beta'(\tau_\mathrm{ext})}{2} \rho_{\mathrm{dis}}$. However, in (\ref{eqn:phase}) the factor $\rme^{\rmi q x_0}$ is required for all $q$, not only in the limit of $|q| \to \infty$. This means, that the value of $\Delta \tau$ given above must be considered only an approximation (for details see section \ref{sec:numerics}). To sum up
\begin{eqnarray}
P_{\mathrm{str}}(\tau) = P_{\mathrm{str}, \tau_\mathrm{ext}=0}(\tau + \Delta \tau), \quad\mathrm{if} \  |\tau| \to 0,
\label{eqn:P_central}
\end{eqnarray}
where $P_{\mathrm{str}, \tau_\mathrm{ext}=0}$ denotes the stress distribution function of the unstressed case, and
\begin{eqnarray}
\Delta \tau \approx Gb \frac{\beta'(\tau_\mathrm{ext})}{2} \rho_{\mathrm{dis}}.
\label{eqn:delta_tau}
\end{eqnarray}

\section{Numerical results}
\label{sec:numerics}

In this section, we present numerically obtained stress distribution functions and compare them with the theoretical predictions discussed above. First, the monodisperse case, for which analytical solution was given in section~\ref{sec:stress}, is considered. Afterwards, real relaxed 2D configurations are investigated.

\subsection{The stress distribution function in monodisperse dipole systems}

In section \ref{sec:monod} the monodisperse dislocation dipole configuration consisting of randomly distributed dipoles having the same slip line distance was introduced and analysed. Each dipole was deformed separately due to the externally applied stress (see figure \ref{fig:dipole_def}). For symmetry reasons the numbers of the dipoles with the four possible dipole angles $45^\circ$, $135^\circ$, $225^\circ$ and $315^\circ$ were taken to be equal.

The reason for analysing the monodisperse dipole system, is the fact that in real relaxed dislocation configurations $d_2^{+-}$ is mainly influenced by the narrow dipoles (see section \ref{sec:corr}). In this model system the $d_2^{+-}$ function can be analytically given [see (\ref{eqn:d_monod}), or (\ref{eqn:d_disp}) if the dipole height has a certain distribution], which permits to obtain an analytical result for the stress distribution function [(\ref{eqn:a_prime_monod}), (\ref{eqn:b_prime_monod}), (\ref{eqn:P_asymp_final}), (\ref{eqn:U_def}), (\ref{eqn:P_central}), and (\ref{eqn:delta_tau})]. In this section this analytical prediction is validated, proving the correctness of the presented deduction.

\begin{figure}[!ht]
\begin{center}
\includegraphics[width=6cm, angle=-90]{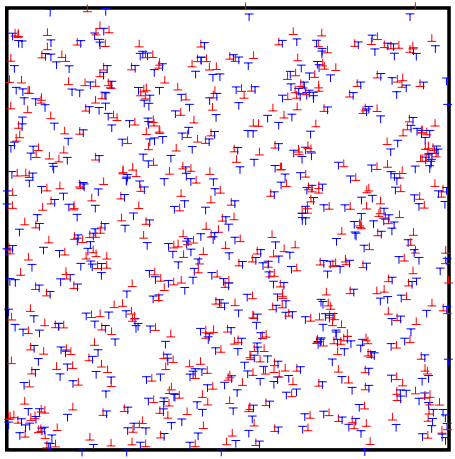}
\end{center}
\caption{\label{fig:monodisperse_conf} A monodisperse dipole configuration of 1024 dislocations under applied external shear stress. The dipole configurations correspond to the equilibrium condition (\ref{eqn:epsilon_origin}).}
\end{figure}

\begin{figure}[!ht]
\begin{center}
\hspace*{-0.5cm}
\includegraphics[width=6cm, angle=-90]{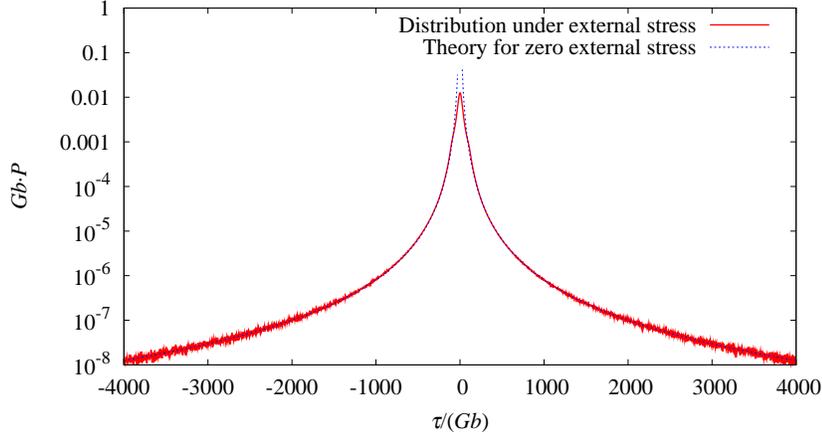}
\end{center}
\caption{\label{fig:monodisperse_4fold_distr} The distribution of internal stresses in a monodisperse dislocation dipole system subjected to external shear stress $\tau_\mathrm{ext} = (5/\sqrt{1024}) \cdot Gb \sqrt{\rho_\mathrm{dis}} \approx 0.156 \cdot Gb \sqrt{\rho_\mathrm{dis}}$. The result was obtained by averaging over 2000 different configurations. Previous result for the tail at zero stress level is also plotted \cite{csikor}.}
\end{figure}

A monodisperse dipole configuration can be easily generated numerically according to its formal definition in section \ref{sec:monod}. An example is seen in figure \ref{fig:monodisperse_conf}. In order to determine the distribution function of internal stresses, the stress values at the grid points of a smooth squared mesh were determined using (\ref{eqn:stress_r}). For the stress field (\ref{eqn:tauind}) was used. (It must be noted that it was numerically verified that if one uses $\tau_\mathrm{ind}$ corresponding to periodic boundary conditions, it does not affect the results presented in this section.) To achieve better statistics, the average of the distribution functions obtained on many different monodisperse configurations consisting of 512 positive and 512 negative dislocations was taken. A typical result can be seen in figure \ref{fig:monodisperse_4fold_distr}.

\begin{figure}[!ht]
\begin{center}
\hspace*{-0.5cm}
\includegraphics[width=6cm, angle=-90]{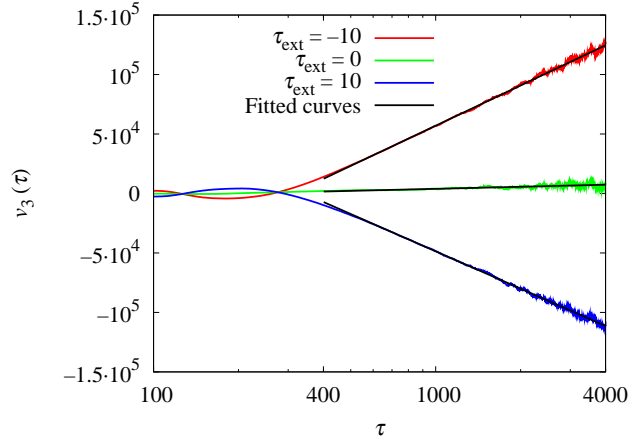}
\end{center}
\caption{\label{fig:monodisperse_4fold_constr_moms} The $v_3$ restricted moments calculated from the distribution functions according to (\ref{eqn:v3}). By plotting $v_3(\tau)$ as a function of $\ln(\tau)$ the resulting linear curve proves the $\tau^{-4}$ like term in the tail of the distribution function [see (\ref{eqn:v3_tail})]. The method was repeated for different applied external stresses $\tau_\mathrm{ext}$ (measured in $Gb\sqrt{\rho_\mathrm{dis}}$ dimensionless units). The results were obtained by averaging over 2000 different configurations.}
\end{figure}

\begin{figure}[!ht]
\begin{center}
\hspace*{-0.5cm}
\includegraphics[width=6cm, angle=-90]{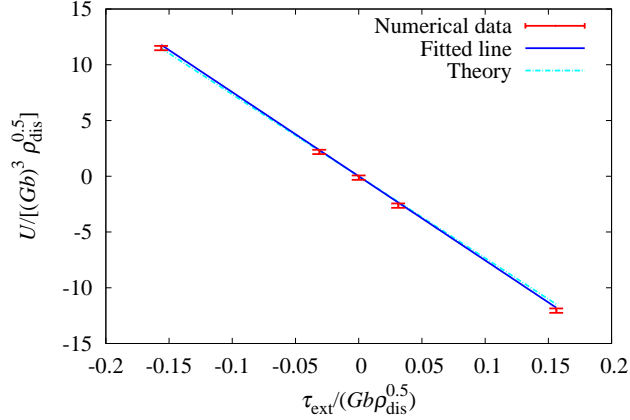}
\end{center}
\caption{\label{fig:monodisperse_4fold_constr_moms_slopes} The values of $U$ obtained by fitting to $v_3(\tau)$ curves in figure \ref{fig:monodisperse_4fold_constr_moms} for different external stresses $\tau_\mathrm{ext}$ [see (\ref{eqn:v3_tail}) and (\ref{eqn:U_def})]. The linear dependence obtained is in agreement with the theory [see (\ref{eqn:a_prime_monod})].}
\end{figure}

At first glance, there is only a little difference between the distribution functions in the stressed and unstressed cases. To prove the existence of the additional $\tau^{-4}$ like term in the tail of the distribution, the $v_3$ third order restricted moment, defined by (\ref{eqn:v3}), was calculated. According to (\ref{eqn:v3_tail}), plotting $v_3(\tau)$ as a function of $\ln(\tau)$ must result in a linear curve. Its slope should be proportional to the external stress. In figure~\ref{fig:monodisperse_4fold_constr_moms} this method is applied for different external stress values. The existence of the $\tau^{-4}$ term is clearly confirmed. The linear dependence of the coefficient $\alpha'(\tau_\mathrm{ext})$ on the external stress $\tau_\mathrm{ext}$ is also verified in figure \ref{fig:monodisperse_4fold_constr_moms_slopes}. By fitting a straight line one obtains
\begin{eqnarray}
U = -2.36 (Gb)^2 \tau_\mathrm{ext}.
\end{eqnarray}
From (\ref{eqn:U_def}) we get
\begin{eqnarray}
\alpha'(\tau_\mathrm{ext}) = -1.57 \frac{\tau_\mathrm{ext}}{Gb}
\end{eqnarray}
meaning there is only about a $2.5\%$ difference between the numerically and theoretically obtained coefficient of the $\tau^{-4}$ tail [see (\ref{eqn:a_prime_monod})].

\begin{figure}[!ht]
\begin{center}
\hspace*{-0.5cm}
\includegraphics[width=6cm, angle=-90]{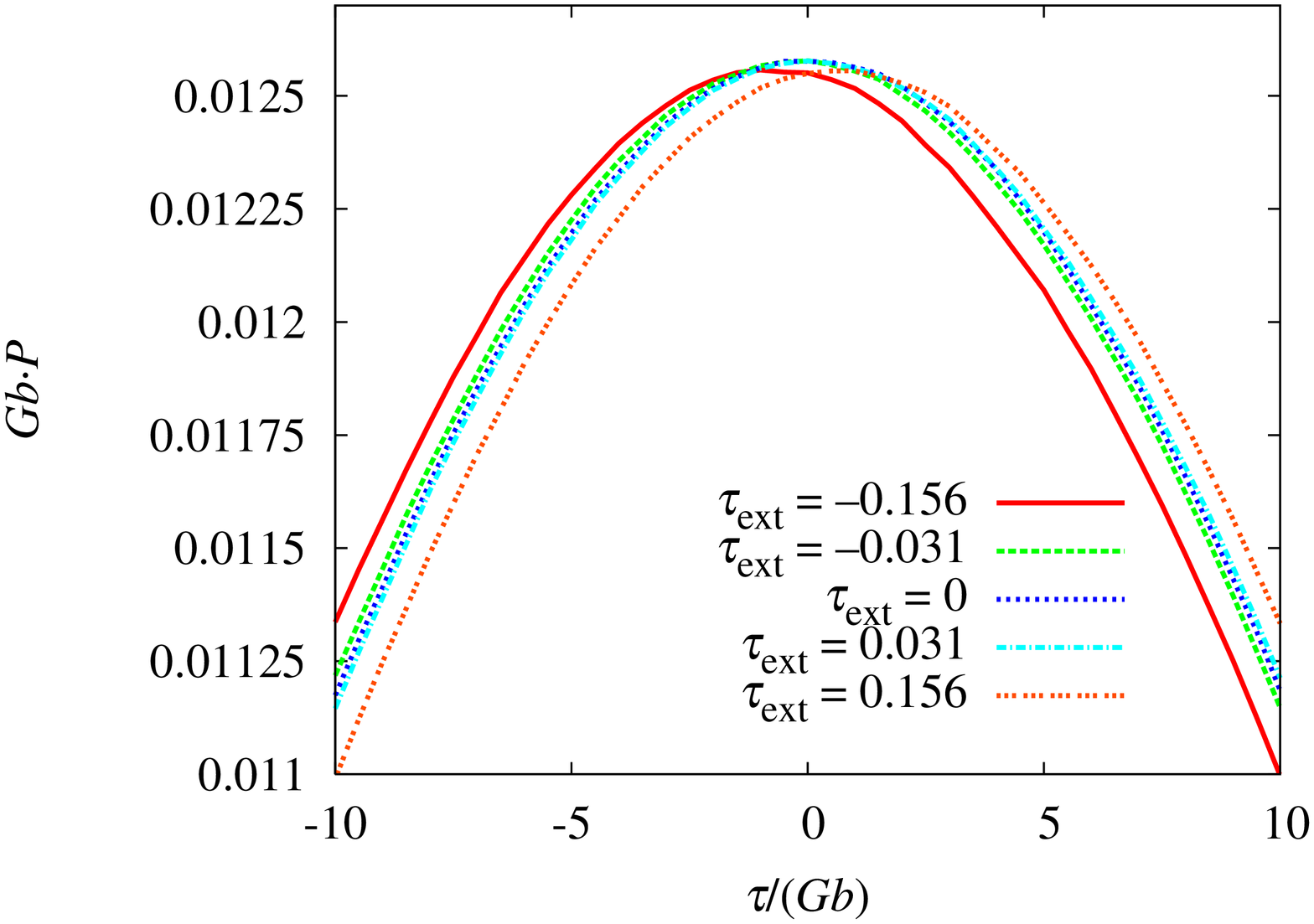}
\end{center}
\caption{\label{fig:monodisperse_4fold_distrib_central} The central part of the stress distribution function of a monodisperse dislocation dipole system under different applied external shear stresses. The result was obtained by averaging over 2000 configurations. (The external stress $\tau_\mathrm{ext}$ is measured in $Gb\sqrt{\rho_\mathrm{dis}}$ dimensionless units.)}
\end{figure}

\begin{figure}[!ht]
\begin{center}
\hspace*{-0.5cm}
\includegraphics[width=6cm, angle=-90]{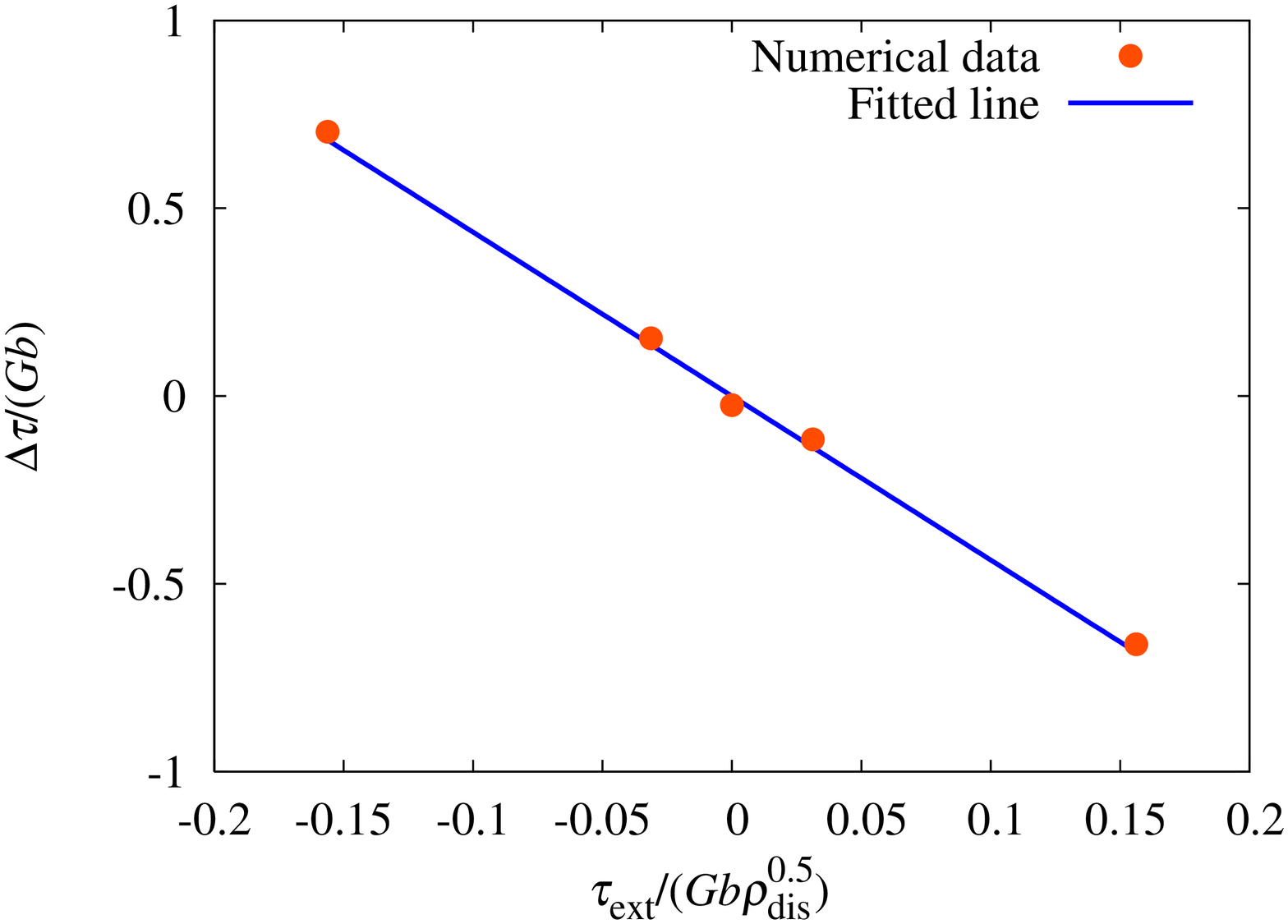}
\end{center}
\caption{\label{fig:monodisperse_4fold_distrib_central_shifts} The shift of the Lorentzian describing the central part of the stress distribution function of a monodisperse dislocation dipole system under different applied external shear stresses. Again, the distribution function is the average of 2000 different realizations.}
\end{figure}

It was derived theoretically that due to the external stress $\tau_\mathrm{ext}$ the central Lorentzian-like part of the stress distribution function is shifted with $\Delta \tau$, which is proportional to $\tau_\mathrm{ext}$ [see (\ref{eqn:P_central}) and (\ref{eqn:delta_tau})]. The shifting can be seen in figure \ref{fig:monodisperse_4fold_distrib_central}. The $\Delta \tau$ values corresponding to different external stresses are plotted in figure \ref{fig:monodisperse_4fold_distrib_central_shifts}. The linear stress dependence of the shifting is clearly fulfilled. Fitting a straight line yields
\begin{eqnarray}
\beta'(\tau_\mathrm{ext}) = -2.67 \frac{\tau_\mathrm{ext}}{Gb} y_0^2
\end{eqnarray}
for $y_0 = 0.32 \cdot \rho_\mathrm{dis}^{-0.5}$. There is about a $20\%$ difference between this numerical result and the theoretical prediction given by (\ref{eqn:b_prime_monod}) and (\ref{eqn:delta_tau}). As we mentioned already in section \ref{sec:central}, (\ref{eqn:delta_tau}) serves only as an estimate for the shifting, therefore, the error of $20\%$ means that numerical results are in a quite good agreement with the theoretical prediction.

\subsection{The stress distribution function in relaxed systems}

In this section we repeat the previous calculations for real relaxed 2D configurations obtained by discrete dislocation dynamics simulations. Initially 64 positive and 64 negative dislocations were placed randomly in a square domain, then they were let to relax under applied external shear stress with the conventional overdamped dynamics \cite{miguel, groma1, miguel3}. To emulate infinite medium, periodic boundary conditions were used (for details about the stress field generated by a dislocation under these conditions see \cite{bako}). In order to reduce computational time, annihilation was introduced for dislocations of opposite signs if their distance became less then $0.02 \rho_\mathrm{dis}^{-0.5}$. This affected in average only the 5\% of starting dislocations. The stress distribution function was determined in the same way described in the previous section. Again, it was found that the distribution function is not influenced by the boundary conditions imposed for the stress. To obtain acceptable results 2000 parallel simulations were carried out and averaging was performed over the stress distribution functions.

\begin{figure}[!ht]
\begin{center}
\hspace*{-0.5cm}
\includegraphics[width=6cm, angle=-90]{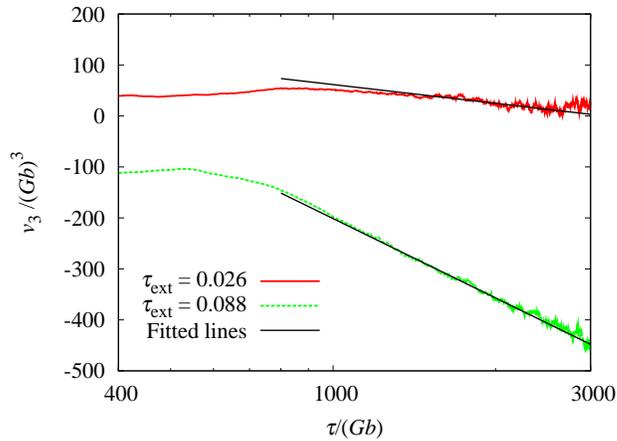}
\end{center}
\caption{\label{fig:relaxed_constr_moms} The $v_3$ restricted moments calculated from the distribution functions according to (\ref{eqn:v3}). By plotting $v_3(\tau)$ as a function of $\ln(\tau)$ the resulting linear curve proves the $\tau^{-4}$ like term in the tail of the distribution function [see (\ref{eqn:v3_tail})]. The method was repeated for two different applied external stresses $\tau_\mathrm{ext}$ (measured in $Gb\sqrt{\rho_\mathrm{dis}}$ dimensionless units). The results were obtained by averaging over 2000 different relaxed configurations.}
\end{figure}

\begin{figure}[!ht]
\begin{center}
\hspace*{-0.5cm}
\includegraphics[width=6cm, angle=-90]{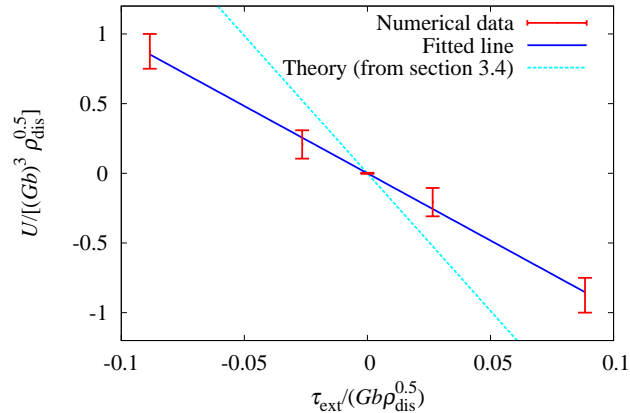}
\end{center}
\caption{\label{fig:relaxed_constr_moms_slopes} The $U$ values obtained by fitting to $v_3(\tau)$ curves in figure \ref{fig:relaxed_constr_moms} for different external stresses [see (\ref{eqn:v3_tail})]. The obtained linear dependence is in agreement with the theory [see (\ref{eqn:a_prime_real2})], however its coefficient is different (the explanation is discussed in the text). For $\tau_\mathrm{ext} < 0$ values the $U_{\tau_\mathrm{ext}} = -U_{-\tau_\mathrm{ext}}$ relation was used (see the text for details).}
\end{figure}

Like in the previous section, in order to prove the asymptote (\ref{eqn:P_asymp_final}), in figure \ref{fig:relaxed_constr_moms} the third order restricted moment $v_3(\tau)$ was plotted against $\ln(\tau)$ [see (\ref{eqn:v3_tail})]. Due to the huge computational demand of the calculations, simulations were only performed with two different external stresses. At $\tau_\mathrm{ext} = 0$ for symmetry reasons $P_\mathrm{str}(\tau) = P_\mathrm{str}(-\tau)$. Therefore $v_3(\tau)$ must vanish. For opposite, non-zero external stresses, again for symmetry reasons, $P_{\mathrm{str}, \tau_\mathrm{ext}}(\tau) = P_{\mathrm{str}, -\tau_\mathrm{ext}}(-\tau)$. This implies that
\begin{eqnarray}
v_{3, \tau_\mathrm{ext}}(\tau) = - v_{3, -\tau_\mathrm{ext}}(\tau).
\label{eqn:v3_symm}
\end{eqnarray}
The values of $U$ were determined according to (\ref{eqn:v3_tail}) and (\ref{eqn:U_def}). The results are plotted in figure \ref{fig:relaxed_constr_moms_slopes} together with the theoretical prediction given by (\ref{eqn:a_prime_real2}). It can be seen that the coefficient $U$ indeed depends linearly on the external stress. By fitting a straight line to the data points we get
\begin{eqnarray}
U = -0.85 (Gb)^2 \tau_\mathrm{ext}.
\end{eqnarray}
According to (\ref{eqn:U_def})
\begin{eqnarray}
\alpha'(\tau_\mathrm{ext}) = -0.57 \frac{\tau_\mathrm{ext}}{Gb}.
\end{eqnarray}
This means that the measured slope of the $U(\tau_\mathrm{ext})$ function is about half of the one calculated from the dipole approximation.

\begin{figure}[!ht]
\psfrag{####ep###}{\hspace*{-0.3cm}$\varepsilon_\mathrm{m}^+(\tau_\mathrm{ext}, y_0)$}
\psfrag{##dx##}{$y_0$}
\begin{center}
\includegraphics[width=5cm]{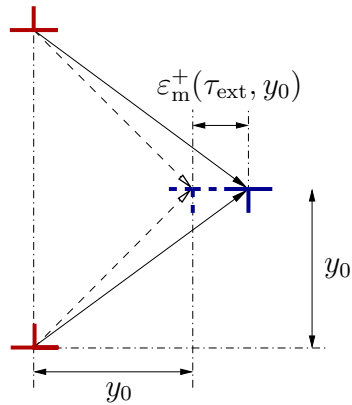}
\caption{\label{fig:multipole_def} The deformation of a certain dislocation multipole under $\tau_\mathrm{ext}$ external shear stress. The change of the $x$ coordinate of the negative dislocation is denoted by $\varepsilon_\mathrm{m}^+(\tau_\mathrm{ext}, y_0)$ which is here half of the similar deformation of a single dipole: $\varepsilon_\mathrm{m}^+(\tau_\mathrm{ext}, y_0) = \varepsilon^+(\tau_\mathrm{ext}, y_0)/2$.}
\end{center}
\end{figure}

The difference observed can be attributed to the fact that in the considerations explained above it was assumed that the relaxed system is mainly built up from small dislocation dipoles deformed only due to the external stress, i.e. the dipole-dipole interaction was neglected. In real systems, however, this is not the case. If one looks at figure \ref{fig:relaxed}(b), beside dipoles, dislocation multipoles can also be observed in a large number. As it is indicated in figure \ref{fig:multipole_def}, a dislocation multipole is always `harder' than a dipole, which formally means $\varepsilon_\mathrm{m}^\pm(\tau_\mathrm{ext}, y_0) < \varepsilon^\pm(\tau_\mathrm{ext}, y_0)$ (see figure \ref{fig:multipole_def} for the definition of $\varepsilon_\mathrm{m}^\pm$). For a small external stress, however, $\varepsilon_\mathrm{m}^\pm(\tau_\mathrm{ext}, y_0) \propto \tau_\mathrm{ext}$ still holds. So, the asymptotic decay of the distribution function remains the same, only the coefficient in (\ref{eqn:a_prime_real2}) is smaller.

\begin{figure}[!ht]
\begin{center}
\hspace*{-0.5cm}
\includegraphics[width=6cm, angle=-90]{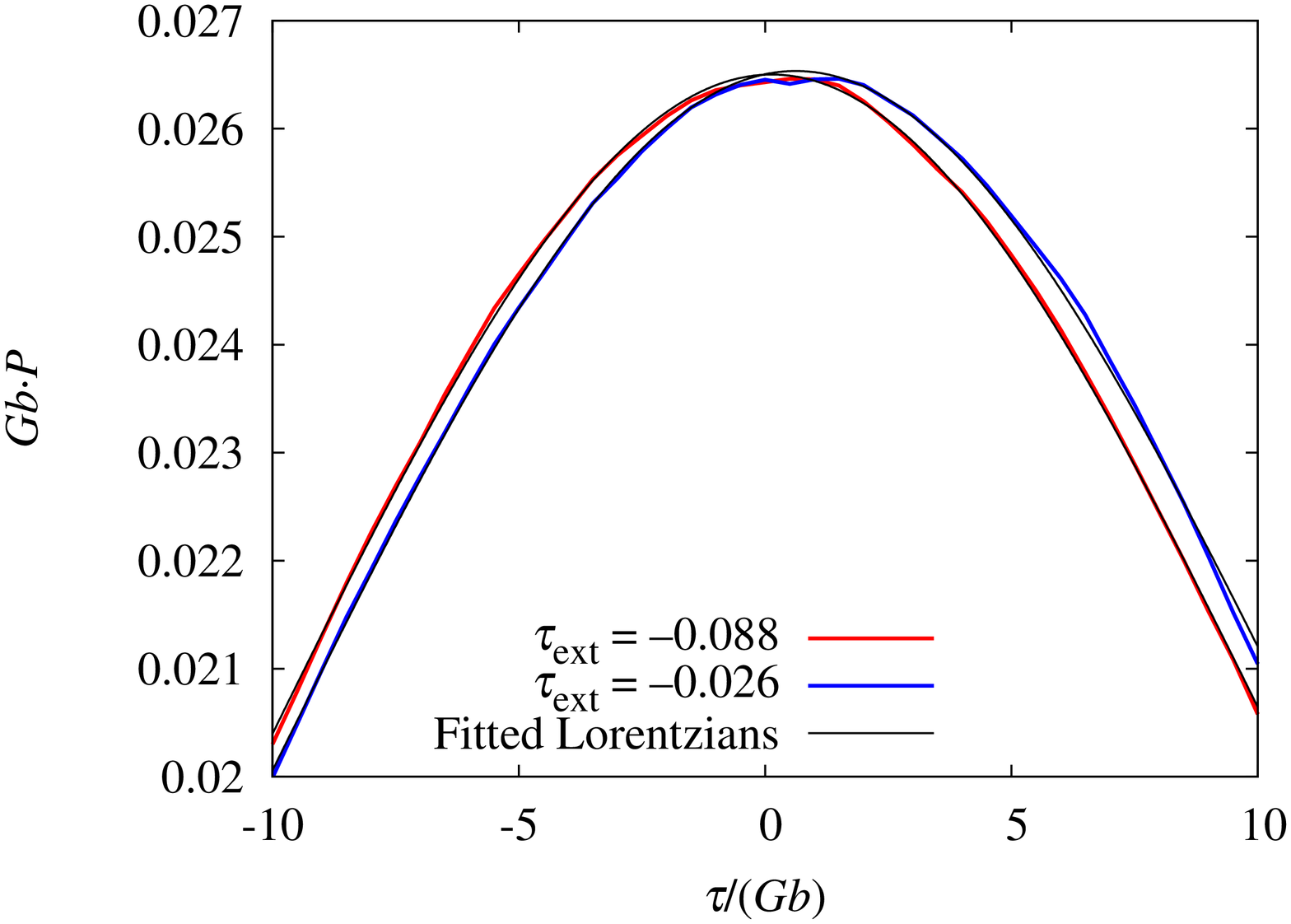}
\end{center}
\caption{\label{fig:relaxed_distrib_central} The central part of the stress distribution function of relaxed dislocation caused by external shear stresses. The result was obtained by averaging over 2000 different configurations.}
\end{figure}

\begin{figure}[!ht]
\begin{center}
\hspace*{-0.5cm}
\includegraphics[width=6cm, angle=-90]{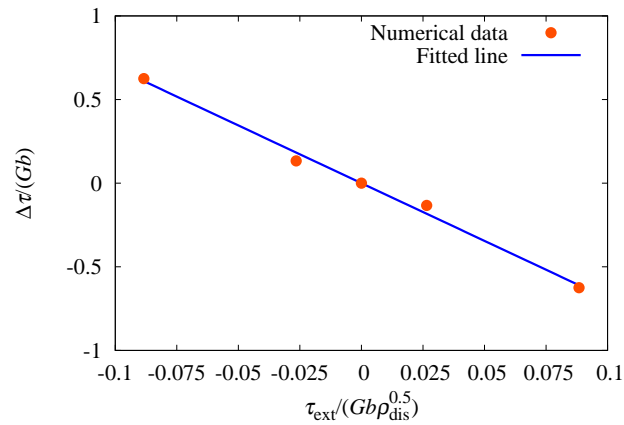}
\end{center}
\caption{\label{fig:relaxed_distrib_central_shifts} The shift of the Lorentzian describing the central part of the stress distribution function of a relaxed dislocation dipole system under different applied external shear stresses.}
\end{figure}

Concerning the shift of the central part of the distribution function (see figure \ref{fig:relaxed_distrib_central}), it can be seen in figure \ref{fig:relaxed_distrib_central_shifts} that the shift $\Delta \tau$ is proportional to the external stress with
\begin{eqnarray}
\beta'(\tau_\mathrm{ext}) = -\frac{1.22}{\rho_\mathrm{dis}} \frac{\tau_\mathrm{ext}}{Gb}.
\end{eqnarray}

\section{Conclusions}

A detailed analysis of the distribution of internal shear stresses in a 2D dislocation system subjected to external shear stress was presented. The results can be summarised as follows:
\begin{enumerate}
\item It was shown theoretically that due to applied stress the initially symmetric stress distribution function becomes asymmetric:
\begin{enumerate}
\item A term proportional to $1/(\tau|\tau|^3)$ is added to the $1/|\tau|^3$ like tail of the distribution function. The coefficient of the extra term is proportional to the stress applied.
\item The central Lorentzian like part of the stress distribution function \cite{csikor} is shifted with a value proportional to the applied stress.
\end{enumerate}
\item For the monodisperse dislocation dipole system the theoretical predictions were proved by the numerical calculation of the stress distribution functions.
\item The distribution function was determined for 2D relaxed dislocation configurations generated by discrete dislocation dynamics simulations at different applied external stresses, too. It was found that one must step beyond the dipole approximation because of presence of dislocation multipoles. However, they only make the material `harder', in the sense that larger external stress is needed to achieve the same change in the distribution function as in monodisperse dipole systems.
\end{enumerate}

\ack

P.\ D.\ I.\ would like to thank F.\ F.\ Csikor for useful discussions. The financial support of the Hungarian Scientific Research Fund (OTKA) under Contract No.\ K 67778 and of the European Community's Human Potential Programme under Contract No.\ NMP3-CT-2006-017105 [DIGIMAT] are also gratefully acknowledged.


\section*{References}

\begin{thebibliography}{00}

\bibitem{miguel} Miguel M.-C, Vespignani A, Zapperi S, Weiss J and Grasso J R, \textit{Intermittent dislocation flow in viscoplastic deformation}, 2001 \textit{Nature} \textbf{410} 667

\bibitem{dimiduk} Dimiduk D M, Woodward C, LeSar R and Uchic M D, \textit{Scale-free intermittent flow in crystal plasticity}, 2006 \textit{Science} \textbf{312} 1188

\bibitem{richeton} Richeton T, Dobron P, Chmelik F, Weiss J and Louchet F, \textit{On the critical character of plasticity in metallic single crystals}, 2006 \textit{Mater. Sci. Eng.} A \textbf{424} 190

\bibitem{csikor_science} Csikor F F, Motz C, Weygand D, Zaiser M and Zapperi S, \textit{Dislocation avalanches, strain bursts, and the problem of plastic forming at the micrometer scale}, 2007 \textit{Science} \textbf{318} 251

\bibitem{sedlacek} Sedl\'a\v{c}ek R, Kratochv\'il J and Werner E, \textit{The importance of being curved: bowing dislocations in a continuum description}, 2003 \textit{Philos. Mag.} \textbf{83} 3735

\bibitem{schwarz} Schwarz C, Sedl\'a\v{c}ek R and Werner E, \textit{Refined short-range interactions in the continuum dislocation-based model of plasticity at the microscale}, 2008 \textit{Acta Mater.} \textbf{56} 341

\bibitem{groma1} Groma I, \textit{Link between the microscopic and mesoscopic length-scale description of the collective behavior of dislocations}, 1997 \textit{Phys. Rev.} B \textbf{56} 5807

\bibitem{groma2} Groma I, Csikor F F and Zaiser M, \textit{Spatial correlations and higher-order gradient terms in a continuum description of dislocation dynamics}, 2003 \textit{Acta Mater.} \textbf{51} 1271

\bibitem{groma3} Groma I, Gy\"orgyi G and Kocsis B, \textit{Debye screening of dislocations}, 2006 \textit{Phys. Rev. Lett.} \textbf{96} 165503

\bibitem{elazab1} El-Azab A, \textit{Statistical mechanics treatment of the evolution of dislocation distributions in single crystals}, 2000 \textit{Phys. Rev.} B \textbf{61} 11956

\bibitem{elazab2} El-Azab A, Deng J and Tang M, \textit{Statistical characterization of dislocation ensembles}, 2007 \textit{Philos. Mag.} \textbf{87} 1201

\bibitem{zaiser1} Zaiser M and Hochrainer T, \textit{Some steps towards a continuum representation of 3D dislocation systems}, 2006 \textit{Scripta Mat.} \textbf{54} 717

\bibitem{hochrainer} Hochrainer T, Zaiser M and Gumbsch P, \textit{A three-dimensional continuum theory of dislocation systems: kinematics and mean-field formulation}, 2007 \textit{Philos. Mag.} \textbf{87} 1261

\bibitem{zaiser2} Zaiser M, Nikitas N, Hochrainer T and Aifantis E C, \textit{Modelling size effects using 3D density-based dislocation dynamics}, 2007 \textit{Philos. Mag.} \textbf{87} 1283

\bibitem{walgraef1} Walgraef D and Aifantis E C, \textit{Dislocation patterning in fatigued metals as a result of dynamical instabilities}, 1985 \textit{J. Appl. Phys.} \textbf{58} 688

\bibitem{walgraef2} Walgraef D and Aifantis E C, \textit{On the formation and stability of dislocation patterns--II. Two-dimensional considerations}, 1985 \textit{Int. J. Eng. Sci.} \textbf{23} 1359

\bibitem{zaisermiguel} Zaiser M, Miguel M.-C and Groma I, \textit{Statistical dynamics of dislocation systems: The influence of dislocation-dislocation correlations}, 2001 \textit{Phys.\ Rev.} B \textbf{64} 224102

\bibitem{deng} Deng J and El-Azab A, \textit{Dislocation pair correlations from dislocation dynamics simulations}, 2007 \textit{J. Computer-Aided Mater. Des.} \textbf{14} 295

\bibitem{csikor_corr} Csikor F F, Groma I, Hochrainer T, Weygand D and Zaiser M, \textit{On the range of 3D dislocation pair correlations}, 2007 \textit{Proc. 11th Int. Symp. on Continuum Models and Discrete Systems (Paris)} (Paris: Mines ParisTech Les Presses) 271

\bibitem{vinogradov} Vinogradov V and Willis J R, \textit{The pair distribution function for an array of screw dislocations}, 2008 \textit{Int. J. Solids Struct.} \textbf{45} 3726

\bibitem{ispanovity} Isp\'anovity P D and Groma I, \textit{Evolution of the correlation functions in two-dimensional dislocation systems}, 2008 \textit{Phys. Rev.} B \textbf{78} 024119

\bibitem{gromabako} Groma I and Bak\'o B, \textit{Probability distribution of internal stresses in parallel straight dislocation systems}, 1998 \textit{Phys.\ Rev.} B \textbf{58} 2969

\bibitem{csikor} Csikor F F and Groma I, \textit{Probability distribution of internal stress in relaxed dislocation systems}, 2004 \textit{Phys.\ Rev.} B \textbf{70} 064106

\bibitem{beato} Beato V, Pietronero L and Zapperi S, \textit{Statistical properties of dislocation mutual interactions}, 2005 \textit{J. Stat. Mech.} P04011

\bibitem{groma4} Groma I, \textit{X-ray line broadening due to an inhomogeneous dislocation distribution}, 1998 \textit{Phys. Rev.} B \textbf{57} 7535

\bibitem{szekely} Sz\'ekely F, Groma I and Lendvai J, \textit{Characterization of self-similar dislocation patterns by x-ray diffraction}, 2000 \textit{Phys. Rev.} B \textbf{62} 3093

\bibitem{borbely} Borb\'ely A and Groma I, \textit{Variance method for the evaluation of particle size and dislocation density from x-ray Bragg peaks}, 2001 \textit{Appl. Phys. Lett.} \textbf{79} 1772

\bibitem{bakogroma} Bak\'o B and Groma I, \textit{Stochastic approach for modeling dislocation patterning}, 1999 \textit{Phys.\ Rev.} B \textbf{60} 122

\bibitem{groma5} Groma I and Bak\'o B, \textit{Dislocation patterning: From micro- to mesoscale description}, 2000 \textit{Phys. Rev. Lett.} \textbf{84} 1487

\bibitem{quinn} Quinn R A and Goree J, \textit{Experimental test of two-dimensional melting through disclination unbinding}, 2001 \textit{Phys. Rev.} E \textbf{64} 051404

\bibitem{nosenko} Nosenko V, Zhdanov S, Ivlev A V, Morfill G, Goree J and Piel A, \textit{Heat transport in a two-dimensional complex (Dusty) plasma at melting conditions}, 2008 \textit{Phys. Rev. Lett.} \textbf{100} 025003

\bibitem{miguel2} Miguel M.-C and Zapperi S, \textit{Tearing transition and plastic flow in superconducting thin films}, 2003 \textit{Nat. Mater.} \textbf{2} 477

\bibitem{murray} Murray C A and Van Winkle D H, \textit{Experimental observation of two-stage melting in a classical two-dimensional screened Coulomb system}, 1987 \textit{Phys. Rev. Lett.} \textbf{58} 1200

\bibitem{schall} Schall P, Cohen I, Weitz D A and Spaepen F, \textit{Visualization of dislocation dynamics in colloidal crystals}, 2004 \textit{Science} \textbf{305} 1944

\bibitem{kader} Abd el Kader A and Earnshaw J C, \textit{Shear-induced changes in two-dimensional foam}, 1999 \textit{Phys. Rev. Lett.} \textbf{82} 2610

\bibitem{markoff} Chandrasekhar S, \textit{Stochastic problems in physics and astronomy}, 1943 \textit{Rev. Mod. Phys.} \textbf{15} 1

\bibitem{rudin} Rudin W 1987 \textit{Real and Complex Analysis} (New York: McGraw Hill) 164

\bibitem{miguel3} Miguel M.-C, Vespignani A, Zaiser M and Zapperi S, \textit{Dislocation jamming and Andrade creep}, 2002 \textit{Phys. Rev. Lett.} \textbf{89} 165501

\bibitem{bako} Bak\'o B, Groma I, Gy\"orgyi G and Zim\'anyi G T, \textit{Dislocation glasses: Aging during relaxation and coarsening}, 2007 \textit{Phys. Rev. Lett.} \textbf{98} 075701

\end{thebibliography}
\end{document}